\documentclass[aps,prb,twocolumn,showpacs,superscriptaddress,citeautoscript]{revtex4}  % for review and submission
\usepackage{graphicx}   % needed for figures
\usepackage{dcolumn}   % needed for some tables
\usepackage{bm}           % for math
\usepackage{amssymb}  % for math
\usepackage{graphicx}
\usepackage{epstopdf, csquotes}
 \usepackage{color,soul}
\usepackage{enumerate}
\usepackage{enumitem}
\usepackage{bm}
\usepackage{amssymb}
\usepackage{amsmath, dsfont}
\usepackage{subfigure, psfrag}
\usepackage{tikz}
\usetikzlibrary{shapes,arrows}
\usepackage{amsmath}
\usepackage{appendix}
\usepackage{multirow}
\usepackage{lipsum, textcase}

\newcommand*{\citen}[1]{%
  \begingroup
    \romannumeral-`\x % remove space at the beginning of \setcitestyle
    \setcitestyle{numbers}%
    \cite{#1}%
  \endgroup   
}

\begin{document}

\title{Anomalous transport in Heisenberg chains with next-nearest neighbor spin-flips}
\title{Anomalous transport in perturbed Heisenberg chains}
\title{Anomalous and regular transport in spin 1/2 chains: AC conductivity}
\author{R.\ J.\ S\'anchez}
\affiliation{Bethe Center for Theoretical Physics, Universit\"{a}t Bonn, Germany}

\author{V.\ K.\ Varma}
\affiliation{Department of Engineering Science and Physics, College of Staten Island, CUNY, Staten Island, NY 10314, USA}
\affiliation{Physics program and Initiative for the Theoretical Sciences, The Graduate Center, CUNY, New York, NY 10016, USA }
\affiliation{Department of Physics and Astronomy, University of Pittsburgh, Pittsburgh, PA 15260, USA}

\author{V.\ Oganesyan}
\affiliation{Department of Engineering Science and Physics, College of Staten Island, CUNY, Staten Island, NY 10314, USA}
\affiliation{Physics program and Initiative for the Theoretical Sciences, The Graduate Center, CUNY, New York, NY 10016, USA }

\date{\today}

\vspace*{-1cm}

\begin{abstract}
We study magnetization transport in anisotropic spin-$1/2$ chains governed by the integrable XXZ model with and without integrability-breaking perturbations at high temperatures ($T\to \infty$) using a hybrid approach that 
combines exact sum-rules with judiciously chosen Ans\"atze.
In the integrable XXZ model we find (i) super-diffusion at the isotropic (Heisenberg) point, with frequency dependent conductivity $ \sigma'(\omega\to 0) \sim |\omega|^{\alpha}$,  where $\alpha=-3/7$ 
in close numerical agreement with recent $t$-DMRG computations; (ii) a continuously drifting exponent from $\alpha=-1^+$ in the XY limit of the model to $\alpha>0$ within the Ising regime; 
and (iii) a diffusion constant saturating in the XY coupling deep in the Ising limit.
We consider two kinds of integrability breaking perturbations --- a simple next-nearest-neighbor spin-flip term ($J_2$) and a three-spin assisted variant ($t_2$), natural in the fermion particle representation of the spin chain. In the first case we discover a remarkable sensitivity of $\sigma'(\omega)$ to the sign of $J_2$, with enhanced low frequency spectral weight and a pronounced upward shift in the magnitude of $\alpha$  for $J_2>0$. Perhaps even more surprising, 
we find sub-diffusion ($\alpha>0$) over a range of $J_2<0$. By contrast, the effects of the \enquote{fermionic} three-spin perturbation are sign symmetric; this perturbation produces a clearly observable hydrodynamic relaxation. At large strength of the integrability breaking term $J_2\to \pm \infty$ the problem is effectively non-interacting (fermions hopping on odd and even sublattices) and we find $\alpha\to -1$ behavior reminiscent of the XY limit of the XXZ chain. Exact diagonalization studies largely corroborate these findings at mid-frequencies.
\end{abstract}
%
%\pacs{05.30.Rt, 05.30.Jp, 67.25.D-, 03.75.Lm}
\maketitle
\section{Background and outline}
Localized deviations from thermal equilibrium in generic strongly correlated many-body systems should relax quickly\citep{ChaikinLubensky}.
However, if the disturbance contains a conserved density 
the corresponding initial imbalance can only relax by spreading slowly (``diffusing") over the entire system as a long-lived collective mode. The characteristic size of such a diffusing distortion scales 
%in one dimension 
as a power law of time $\langle (\Delta x)^2 \rangle \sim t^{1-\alpha}$, where $\alpha = 0$ for ordinary diffusion. The presence of additional %local (or quasilocal) 
conserved quantities can either slow down or speed up the spread, e.g.~in Anderson or many-body localized systems $\langle (\Delta x)^2 \rangle \sim t^0$, 
while in the presence of ballistically propagating quasi-particles we expect $\langle (\Delta x)^2 \rangle \sim t^2$, i.e.~\emph{effectively} $\alpha=1$ and $\alpha=-1$, respectively. 
A complimentary view of relaxation processes is suggested by writing the linear response conductivity in terms of current \emph{fluctuations} \cite{Mahan}
\begin{equation}
\sigma'(\omega)=\frac{1 - \textrm{e}^{-\beta \omega}}{L \, \omega} \int_{0}^\infty dt \, \cos(\omega t) \langle \hat j (t) \hat j (0)\rangle,
\label{eq:fluctKubo}
\end{equation}
where $\hat j$ is the macroscopic current (see Eq.~\eqref{eq:current} below), $\beta$ is the inverse temperature, $L$ is the system size and $\langle\ldots\rangle$ denotes equilibrium average.
The conductivity can include both a regular dissipative component as well as a \enquote{ballistic} zero-frequency component
%We may consider explicitly current-carrying initial conditions and identify ballistic zero-frequency non-dissipative behavior with late time persistent current 
\begin{equation}
\sigma'(\omega) = D \, \delta(\omega) + \sigma_{\mbox{\footnotesize reg}}(\omega),
\end{equation}

\noindent 
with $D$ commonly referred to as the \enquote{Drude weight} and given by the long-time asymptotic value of the current auto-correlation
\begin{equation}
\frac{\pi \beta^{-1}}{L}\lim_{t\rightarrow \infty} %\int_0^{\tau} dt 
\langle \hat j (t) \hat j (0)\rangle = D \ge 0.
\label{eq:drude-longtime}
\end{equation}

There are several well studied examples of so-called integrable lattice models with $D>0$~\cite{PhysRevLett.74.972, PhysRevB.53.983, PhysRevB.55.11029}, 
typically both at zero and finite temperatures\cite{continuum}. 
Our main focus is the comparatively less well understood regular part of the conductivity at finite frequencies, which is usually finite in interacting lattice models. The presence of additional conserved quantities 
is expected to induce slow non-exponential decay of current fluctuations in time, leading to interesting super- or sub-diffusive phenomena. 
The latter necessarily follows a very rapid initial decay of the current that overshoots zero, and so the final decay contains an overall negative sign --- suppression of conductivity at low frequency necessarily 
requires large cancellations between long and short times in Eq.~\eqref{eq:fluctKubo}. 

Our methodology, as summarized in Section~\ref{sec:definitions}, makes no explicit use of the detailed dynamical structure of the problems we study, the choice of models and Ans\"atze does. Our results may be roughly divided into two categories: identification of qualitatively new hydrodynamic behaviors (Sections~\ref{sec: analytic_results} and~\ref{sec: NearlyFree}), and quantitative understanding of relatively conventional hydrodynamics (Sections~\ref{sec:Isingatomic} and~\ref{sec:nonlinhydro}). 
Section~\ref{sec: analytic_results} examines the vicinity of the integrable Heisenberg point of the XXZ chain. The latter appears perched near the boundary between super- and sub-diffusive tendencies. 
The dominant effect of weak nearest-neighbor spin-flip perturbation appears to be to traverse this boundary.
Section~\ref{sec:Isingatomic} constructs a microscopic description of the two-component dynamic response in the so-called \enquote{classical} Ising regime (also referred to as the \enquote{atomic} limit).
Section~\ref{sec: NearlyFree} focuses on weakly interacting fermion regimes in the generalized XXZ Hamiltonian, exhibiting logarithmic (!) current relaxation 
$$\langle \hat j(t) \hat j(0)\rangle \sim -\log t,$$ 
over a broad dynamical regime.
Section~\ref{sec:nonlinhydro} demonstrates a highly accurate analytic approximation to the non-linearity induced long-time tails in strongly ergodic chains, realized with the help of the \enquote{fermionic} $t_2$ perturbation. 
Implications for inelastic scattering experiments are discussed in Section~\ref{sec:experiments}, with conclusions summarized in Section~\ref{sec:conclusion}.  Several appendices treat important further details and nuances.
\section{Definitions,
general considerations, relevant prior results}
\label{sec:definitions}
\subsection{Hamiltonians, currents, global symmetries}
We consider spin chains of length $L$ with periodic boundary conditions (possibly with a twist, as explained below) and dynamics governed by the Hamiltonian $\hat H_0$ of the nearest neighbor (nn) XXZ model, perturbed by one of two types of next-nearest neighbor (nnn) spin-flip terms ($\delta \hat H_1$ and $\delta \hat H_2$)
\begin{eqnarray}
\hat H_0 &=&  \sum_{j}^L \frac{J_1}{2} (\hat{S}_j^{+}\hat{S}^{-}_{j+1} + \textrm{h.c}) + \Delta \sum_{j}^L \hat{S}^{z}_j\hat{S}^{z}_{j+1},
%\raggedleft
\label{eq: hamiltonian}
\\
\delta \hat H_1 &=&  \sum_{j}^L \frac{J_2}{2} (\hat{S}_j^{+}\hat{S}^{-}_{j+2} + \textrm{h.c}),
%\raggedleft
\label{eq: hamiltonianNNN}
\\
\delta \hat H_2 &=&  \sum_{j}^L \frac{t_2}{2} (\hat{S}_j^{+}\hat{S}_{j+1}^{z}\hat{S}^{-}_{j+2} + \textrm{h.c}),
%\raggedleft
\label{eq: hamiltonianNNNf}
\end{eqnarray}
\noindent
where the raising and lowering spin operators ($\hat{S}_j^{+}\hat{S}^{-}_{j+i} + \textrm{h.c}$) flip pairs of spins at sites $j$ and $j+i$ with amplitude $J_i$, and the Ising coupling is 
$\Delta>0$. 
The $t_2$ 3-spin nnn perturbation appearing in $\delta\hat  H_2$ is natural in the spinless fermion representation, where it corresponds to nnn fermion hopping.%is quadratic in the fermionic operators. 
 
The Hamiltonian $\hat H_0$ %in Eq.~\eqref{eq: hamiltonian} 
is invariant under spin rotations about the $z$-axis $\hat m_z$, lattice translations $\hat T$,  spin-inversion $\hat Z$ and space-reflection, or parity, $\hat P$. 
The spin-inversion operator, with eigenvalues $z=\pm 1$,  is defined such as
\begin{equation}
\hat Z \hat S_j^{\pm} \hat Z = \hat S_j^{\mp}, \quad \hat Z \hat S_j^{z} \hat Z = -\hat S_j^{z}.
\end{equation}

Likewise the parity operator, with eigenvalues $p=\pm 1$, is defined as 
\begin{equation}
\hat P \hat S_j^{\gamma} \hat P = \hat S_{L-j}^{\gamma}, \quad \mbox{with} \, \gamma = \pm,\,  z.
\end{equation}

Thus, in addition to the total energy, the total magnetization along the $z$-axis ``$m_z$", the total crystal momentum ``$K$", the $z$-parity ``$z$" and parity ``$p$" are global conserved quantities of the system. Note however that although $\delta \hat H_1$ is also invariant under this set of operations, $\delta \hat H_2$ does not commute with $\hat Z$.

We define the total spin current operator for the full model ($\hat H_0 + \delta \hat H_1+\delta \hat H_2$) as $\hat j^z = \sum_j^L \hat j_l^z$, where the local current operator $\hat j_l^z$ satisfies the continuity equation 
%We now define the spin current operator, $\hat j^z = \sum_j^L \hat j_l^z$, we shall use in our discussion of spin transport. Since $m_z$ is conserved, the local current operator $\hat j_l^z$ satisfies the continuity equation 
\begin{equation}
\dot{S}_l^z= -i [\hat H, \hat S_l^z]\equiv\hat j_l^z - \hat j_{l-1}^z,
\label{eq: continuity-equation}
\end{equation}
which then gives
\begin{eqnarray}
\hat{j^z} &=&
i\sum_{j}\left(\frac{J_1}{2} \, \hat{S}_j^{+}\hat{S}^{-}_{j+1} + J_2 \, \hat{S}_j^{+}\hat{S}^{-}_{j+2}
+ t_2 \, \hat{S}_j^{+}\hat{S}^{z}_{j+1}\hat{S}^{-}_{j+2}\right) \nonumber \\
&+& \textrm{h.c.}
\label{eq:current}
\end{eqnarray}

Note this operator is odd under both spin-inversion (in the absence of $\delta \hat H_2$) and parity, i.e. 
\begin{equation}
\hat{O} \, \hat{j^z} \, \hat{O} = -\hat{j^z}, \quad \mbox{for} \, \, \,  \hat{O} = \hat Z, \, \hat P.
\label{eq: j_oddness}
\end{equation}

%\vspace{0.2cm}
%{\it Current operator ---} 
\subsection{Linear response conductivity, series expansion and Ans\"atze}
%\vspace{0.2cm}
%{\it When is transport ballistic? ---} 
Following standard derivations we will study the linear response (Kubo) conductivity
\begin{eqnarray}
\beta^{-1}\sigma'(\omega)&=&\frac{\pi}{LZ}\sum_{m,n} \delta(\omega-E_m+E_n)|j^z_{mn}|^2,
\label{eq:Kubo}
\end{eqnarray}
here simplified to its high temperature limiting form, where $Z$ is the partition function of the model and $j^z_{mn}$ labels the matrix elements of the current operator. 

In averaging current auto-correlations we have choices of ensembles, e.g.~averaging over states with fixed total magnetization or over all magnetization sectors, and of  boundary conditions, e.g.~ open vs. periodic.
As with thermodynamics, we expect the finite frequency conductivity to be insensitive to these choices (even as $\omega\to 0$), but only in the thermodynamic limit. The choices we make in this work are guided by the types of analytic and numerical tools at our disposal. 
The most natural ensemble choice in the context of analytic moment expansions (see next paragraph) is the infinite temperature unbiased \enquote{grand-canonical} average over all magnetization sectors, 
which is what we will use in the main body of the paper. Since the entropy density peaks near zero magnetization, this averaging weighs contributions from $m_z\approx 0$ states the highest. 
Boundary condition choices, on the other hand, are guided by numerical exact diagonalization considerations. 
In particular, residual symmetries, e.g.~parity and spin-inversion, can slow down the convergence to the thermodynamic limit\cite{SanchezVarma}.
Periodic boundary conditions \emph{with} an irrational flux threading the ring is the standard way to remove such transient (see also Appendix~\ref{app: FSEfluxing}). 
Free boundary conditions can in principle play the same role, however, these tend to induce very strong edge effects, especially in the strongly interacting regime of interest here\cite{PhysRevB.77.161101}. 
Thus in the bulk of the paper we adhere to \enquote{grand canonical} averaging and irrationally twisted boundary conditions, but include additional results of our explorations of ensembles and boundary conditions in Appendix~\ref{app: FSEfluxing}. 

Frequency moments of the conductivity and short-time series expansion of the current auto-correlation function are related via 
%\begin{equation}
%\langle j(t)j(0)\rangle /L \equiv \sum_{n=0}^\infty \mu_n \, t^{2n}, \hspace{2em}\mu_n=\int_{-\infty}^\infty \frac{d \omega}{2\pi} \omega^{2n} \beta^{-1}\sigma(\omega)
%\label{eq:shorttime}
%\end{equation}
\begin{equation}
\frac{1}{L} \, \langle \, \hat j^z(t) \hat j^z(0)\rangle \equiv \sum_{n=0}^\infty \frac{\mu_n}{(2n)!} \, t^{2n}, 
\label{eq:shorttime}
\end{equation}
and 
\begin{equation}
\mu_n=\int_{-\infty}^\infty \frac{d \omega}{2\pi} \omega^{2n} \sigma'(\omega).
\label{eq:momenteq1}
\end{equation}

The coefficients $\mu_n$ are nested commutators of the current operator with the Hamiltonian $\hat H$ describing the system, e.g.~$\mu_0=\langle \hat j^z \hat j^z \rangle/L$ and $\mu_1=\langle [[\hat j^z, \hat H], \hat H] \, \hat j^z\rangle/L$. 
Odd terms are necessarily absent in equilibrium.
In practice it is often useful to take advantage of the continuity relationship to obtain the $\mu_n$'s from the small $q$ dependence of the \enquote{density-density} correlator, i.e.~using the identity 
\begin{equation}
\langle \hat j^z(t) \hat j^z(0)\rangle = \frac{\partial^2}{q^2\partial t^2}\langle \hat S^z(q,t) \hat S^z (-q,0)\rangle \biggr |_{q\to 0}.
\label{eq:eom}
\end{equation}

Working with the density means we can start the expansion with a very simple local term, e.g.~a single $\hat S^z$ operator at the origin
%center (i.e.~within the bulk) 
of the spin chain, and take maximum advantage of the locality of the Hamiltonian, 
as subsequent commutators grow the spatial extend of the series linearly. The downside is that we need an extra term in the series because of the time derivatives. 
More explicitly, we use Eq.~\eqref{eq:eom} to express the $\mu_n$'s as thermal expectation values, {\it within the grand canonical ensemble}, of the product of a $(2n+2)$-fold commutators with a spin operator
\begin{equation}
\label{eq: nestedSz}
\mu_n = 
%\sum_{(j-j')}
\lim_{q\rightarrow 0}\frac{1}{q^2 %L
}\sum_{j%,j'
}\textrm{e}^{-i q j}\langle [...[\, [\hat S_j^z, \hat H], \hat H], ..., \hat H] \hat S_{0%j'
}^z \rangle. 
\end{equation}

Short-time series expansions are similar to but less well studied than the high-temperature series expansion known from classical statistical mechanics.  Given a long series there exist several inexact but useful techniques to try and reconstruct the behavior of the function, e.g.~Pad\'e approximants. This is ongoing work on which we hope to report in the near future. Here, instead, we adopt a simpler Ansatz-based approach, where several qualitative features are not computed but rather inserted by fiat into a relatively simple functional form, motivated on prior numerical or analytic results as well as general considerations, e.g.~$\sigma(\omega)\propto |\omega|^\alpha \exp(-a \, \omega^2)$, whose parameters can be computed by requiring it to fulfill a few low-order moments. If the results are relatively accurate somewhere in parameter space, we expect them to continue to be accurate in the immediate vicinity, i.e.~we can take (somewhat) seriously the dependence of the Ansatz's parameters on the coupling constants in the Hamiltonian.

\subsection{Integrability vs. ballisticity}
In the nonintegrable (finite $t_2$ or $J_2$) cases the Drude weight is expected to vanish in the thermodynamic limit --- exponentially with system size, with a $J_2$($t_2$)-dependent exponent, and indeed numeric simulations support this expectation\citep{PhysRevB.53.983}. 
Finite $t_2$ gives rise to nonlinear hydrodynamics at high-temperatures\cite{PhysRevB.73.035113}, whereas positive $J_2$ yields frustration-enhanced transport at low-temperatures\cite{PhysRevA.92.013618}.

On the other hand, at $J_2=t_2=0$ the model is integrable and has a macroscopic number of local conserved quantities $\{\hat Q_m\}$, some of which may have a finite overlap with the spin current operator (i.e.~$\langle \hat  j^z \hat Q_m  \rangle \neq 0$) thereby preventing current auto-correlations from completely decaying. This can be seen clearly by examining the current auto-correlation function in time; the Drude weight is given by the residual non-decaying component of initial currents (see Eq.~\eqref{eq:drude-longtime} above). Thus provided the current operator has components along any of the $\hat Q_m$'s, the Drude weight will be finite and transport will be ballistic.

For Eq.~\eqref{eq: hamiltonian} all these local conserved quantities are even under spin-inversion\cite{PhysRevB.55.11029}, whereas $\hat j^z$ is odd, and consequently the thermal expectation value 
$\langle \hat j^z \hat Q_m \rangle$ vanishes. 
However, if spin inversion symmetry is broken by e.g.~an external magnetic field, a finite magnetization density or a magnetic flux --- the latter only for finite spin rings, the expectation value 
$\langle \hat j^z \hat Q_m \rangle$ need no vanish. 
In fact, using an inequality due to Mazur\citep{Mazur_ineq} it has been shown\citep{PhysRevB.55.11029} that in the high-temperature limit, and for finite magnetization densities, the Drude weight has the 
finite lower bound 

\begin{equation}
D \ge \frac{\beta}{2} \, \frac{8 \, \Delta^2 \,  m_z^2 \, (1/4-m_z^2)}{1+8 \, \Delta^2(1/4+ m_z^2)}.
\label{eq: Mazur-Zotos}
\end{equation}

 Eq.~\eqref{eq: Mazur-Zotos} accounts for the overlap of $\hat j^z$ with the conserved local energy current\citep{PhysRevB.55.11029}, and entails the XXZ model displays ballistic transport for $\Delta>0$ and finite $m_z$. 

At exactly zero magnetization Eq.~\eqref{eq: Mazur-Zotos} is inconclusive and one needs to distinguish between three regimes with different transport properties. For the commensurate anisotropies $\Delta = \cos(\pi M/N)<1$, with $N, M \in \mathbb{Z}$ coprimes and $N>M$ there exists a one-parameter 
family of quasilocal exact conserved quantities, which includes operators of odd $z$-parity having a finite overlap with $\hat j^z$. 
This family of quasilocal operators, together with Mazur's inequality, yields a finite lower bound for the Drude weight and entails ballistic transport for $\Delta < 1$ \citep{PhysRevLett.106.217206, PhysRevLett.111.057203}.
In connection with these results, we have recently conjectured that the degenerate states due to the additional quantum symmetry (partially identified in Ref.~\citen{DeguchiFabriciusMcCoy}) at these commensurate anisotropies correspond to the thermodynamic current-carrying states responsible for ballisticity, and saturate the lower bound\cite{SanchezVarma}.
%
%The role of symmetries and ensembles on the numeric computation of the Drude weight in this limit has been expounded upon by two of us recently, which led to a conjecture on the identification of the thermodynamic current-carrying states\cite{SanchezVarma}.
%
At $\Delta = 1$ the global $SU(2)$ symmetry of the model can be used to construct a rigorous upper bound for the Drude weight. 
This upper bound vanishes with the magnetization, thus implying the absence of ballistic transport at the isotropic point ($\Delta = 1$)\citep{PhysRevB.92.165133, CarmeloProsen_UpperBoundXXXmodelGranCan}. 
Finally, for $\Delta > 1$ the Drude weight is known to vanish, as follows from the spin-reversal invariance of the macrostates sustained by the system in this limit~\citep{PhysRevLett.119.020602}, and as confirmed by numeric simulations~\citep{PhysRevB.53.983, PhysRevB.68.134436}. 

Summarizing, transport at $J_2=t_2=0$ is dominated at late times (low frequencies) by the ballistic (Drude peak) response. The Drude weight vanishes for anisotropies $\Delta \ge 1$ \emph{and} zero magnetization density, where some (likely weaker and/or anomalous) form of diffusion has been proposed and observed in recent works.
%densities is {\it nonballistic}, i.e.~it is dominated by the regular part of the Kubo conductivity.
%In what follows we attempt to estimate the transport rates characterizing the longitudinal spin damping for a system described by Eq.~\eqref{eq: hamiltonian}, both in presence and absence of the perturbations $\delta \hat H_1$ and $\delta \hat H_2$, in the disordered high-temperature limit. 
%
%In so doing we systematically account for such anomalous effects.
%{\it when the Drude weight is expected to vanish}, that is, (i) for finite $J_2$ and $\Delta \in \mathbb{R}$; and (ii) in the integrable case ($J_2=0$) for $\Delta \ge 1$, in the sector of vanishing magnetization. 
\section{Anomalous relaxation with weak or no breaking of integrability}
\label{sec: analytic_results}
We start by setting $t_2=0$ and displaying the first three non-zero conductivity moments 
\begin{eqnarray}
\label{eq:m1}
\mu_0 & = &   \frac{1}{8}J_1^2+\frac{1}{2}J_2^2 , \\ \nonumber \\ 
\label{eq:m2}
\mu_1 & = &  \frac{1}{16}J_1^2\Delta^2+\frac{1}{4}J_2^2\Delta^2 - \frac{3}{8} J_1^2J_2\Delta+\frac{5}{8}J_1^2J_2^2, \\ \nonumber \\ 
\label{eq:m3}
\mu_2 & = &  \frac{1}{16}J_1^2\Delta^4+\frac{1}{4}J_2^2\Delta^4 + \frac{5}{64}J_1^4\Delta^2+\frac{3}{8}J_2^4\Delta^2   \nonumber \\
&  &   + \, \frac{5}{2} J_1^2J_2^2\Delta^2  -\frac{11}{16}J_1^2J_2\Delta^3 -\frac{5}{8}J_1^4J_2\Delta \nonumber \\
&   &  - \, \frac{27}{32}J_1^2J_2^3\Delta +\frac{7}{8}J_1^2J_2^4  +\frac{85}{64}J_1^4J_2^2.
\raggedleft
\end{eqnarray}

The information about $\sigma(\omega)$ we aim to glimpse is the overall integrated weight, the low-frequency anomaly and the high-frequency cutoff. This leaves a lot of freedom which we explored somewhat but in the end settled on the following simple functional form, whose high-frequency envelope is taken to be Gaussian
\cite{ansatz1f, ansatz2f}
\begin{equation}
\beta^{-1}\sigma'(\omega,T) = \mathcal{C} \, \frac{\omega_0^{-1-\alpha} }{\Gamma\left(\frac{1+\alpha}{2}\right)} \, | \omega |^{\alpha} \exp\left(-\Bigr |\frac{\omega}{\omega_0} \Bigr|^2 \right),
\label{eq: ansatz}
\end{equation}
where the parameters $\alpha$ and $\omega_0$ are to be determined in terms of the moments Eqs.~(\ref{eq:m1} -- \ref{eq:m3}), $\Gamma(x)$ is the gamma function, 
and the amplitude $\mathcal{C}$ is fixed by the first sum rule to be $\mathcal{C} = 2\pi \mu_0$.  Combining Eq.~\eqref{eq:momenteq1} with Eqs.~(\ref{eq:m1} -- \ref{eq: ansatz}) we obtain

\begin{eqnarray}
\frac{\mu_n}{\mu_0} & = &\omega_0^{2n-2} \frac{\Gamma\left(\frac{2n-1+\alpha}{2}\right)}{\Gamma\left(\frac{1+\alpha}{2}\right)},
\label{eq: moment-equations}
\\
\alpha & = &-1+\frac{2\mu_1^2}{\mu_0\mu_2-\mu_1^2},
\label{eq: dynamicalexponent}
\\
\omega_0^2 & =&\frac{\mu_2}{\mu_1}-\frac{\mu_1}{\mu_0}.
\label{eq:collision-time}
\end{eqnarray}
\begin{figure*}[ttp!]
\centering 
\includegraphics[width=18.2cm]{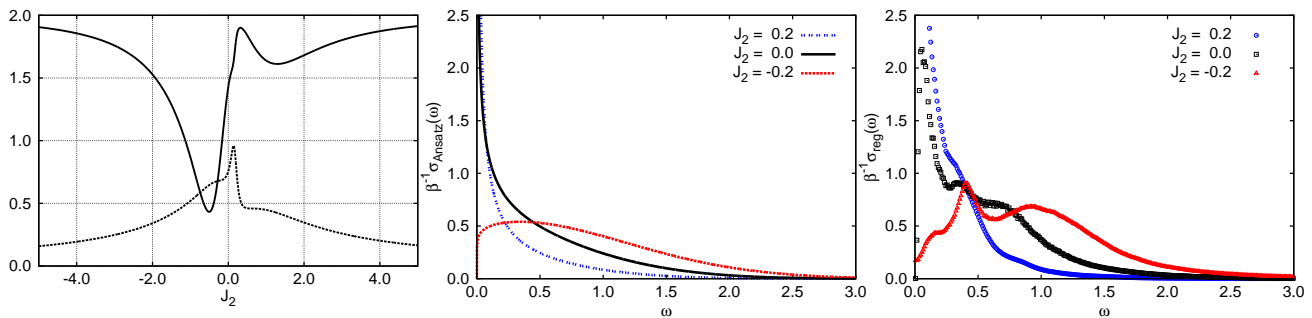}
\caption{{\it Left}: dynamical exponent (full line) characterizing the longitudinal spin damping $\mbox{lim}_{t\rightarrow \infty} \langle (\Delta x)^2 \rangle \sim t^{1-\alpha}$ for $\Delta=1$ as a function of $J_2$. 
The range of $J_2$ values for which $\alpha>0$ makes the system subdiffusive. 
The dashed line indicates the variation of $\omega_0^{-1}$ with $J_2$, also showing the asymmetry with the sign of $J_2$.
{\it Center}: frequency dependence of the (normalized) conductivity Ansatz, for $\Delta=1$ and three values of $J_2$. Note the asymmetry for the transport rates with respect to the sign of $J_2$.  
{\it Right}: normalized Kubo conductivity for a 20-site spin ring, computed within the canonical ensemble for states with zero total magnetization, and the same set of parameters as for the memory function (uniform binning, width = 0.01). All the given results are in units of $J_1$.}
\label{fig: DiffusionExponent}
\end{figure*}

Provided the overall shape of the true AC conductivity is similar to that assumed in the Ansatz,
this crude approach can yield useful insights and predict interesting trends in the dependence on Hamiltonian parameters. In especially fortuitous cases it might turn out to be quantitatively correct, as appears to be the case for the Heisenberg chain, as described below. We shall examine two particular paths through the two-dimensional parameter space --- the integrable XXZ chains ($J_2=t_2=0$) and $J_2$-perturbed non-integrable Heisenberg point ($J_1=\Delta=1$).  At least near the Heisenberg point, we can confirm numerically (see Fig.~\ref{fig: DiffusionExponent}) that the true conductivity and its approximation are similar.

\subsection{The integrable XXZ line, $J_2=0$}
In treating the integrable XXZ model we must differentiate between the Ising ($\Delta>J_1=1$) and the XY ($\Delta<J_1=1$)
regimes\cite{gapped} --- as already mentioned in Section II, the model has a finite Drude weight on the XY side\cite{ProsenGapless} which contributes to the leading frequency moment, while on the Ising side the Drude weight vanishes in the only statistically significant magnetization sector ($m_z=0$). 

\paragraph{XY regime ($\Delta<1$).} If we for simplicity approximate the Drude weight with the Thermodynamic Bethe Ansatz result\cite{Zotos, Benz} 
\begin{equation}
\beta^{-1}D = 2\pi\frac{\gamma - \sin{(2\gamma)/2}}{16\gamma},
\label{eq: ThermodynamicBetheAnsatz}
\end{equation}
with $\gamma = \cos^{-1}\Delta$, and set the leading moment {\it of the regular part} of the conductivity to $\mu_0- D/\pi$ in Eq.~\eqref{eq: dynamicalexponent} we obtain \cite{artefact}
\begin{equation}
\label{eq: exponent_gapless}
 \alpha =\frac{8 \Delta^2 \cos^{-1} \Delta}{(5+4 \Delta^2) \sin (2\cos^{-1} \Delta)-4\Delta^2\cos^{-1}\Delta}-1.
% \nonumber \\   \nonumber \\ 
% &    &
\end{equation}

This result implies the low-frequency spin conductivity displays superdiffusive behavior alongside ballistic relaxation that stems from the Drude peak 
\begin{equation}
\beta^{-1} \sigma'(\omega) \sim D \, \delta(\omega) + \omega^{\alpha},  
\end{equation}

\noindent 
with the exponent $-1 < \alpha < -3/7$.

The seemingly important subtraction of the finite Drude weight produces very mild corrections in $\alpha$ and $\omega_0$. In particular, in the limit $\Delta\to 0$ the leading behavior is governed by the parametric smallness of $\mu_1^2\sim \Delta^4$ vs. $\mu_0 \mu_2 \sim \Delta^3$ even after the subtraction (and $\mu_0 \mu_2 \sim \Delta^2$ without it). 

\paragraph{Ising regime and Heisenberg point ($\Delta\geq 1$).}
Away from the XY regime the Drude weight vanishes and we obtain simpler looking formulae
\begin{eqnarray}
\label{eq: dynamicalexponent_II}
 \alpha &= &\frac{2 \Delta^2-5}{2 \Delta^2+5},
\\ \nonumber \\ 
 \label{eq: halfwidth}
\omega_0 &=& \frac{\sqrt{2\Delta^2 + 5}}{2},
\end{eqnarray}
amounting to a continuous redistribution of spectral weight away from the low frequency divergence to high frequency tails, a physically sensible result. More intriguingly, the exponent crosses zero at $\Delta=\sqrt{5/2}\approx 1.6$ with the conductivity becoming subdiffusive, i.e.~decreasing at lower frequencies for larger $\Delta$. Subdiffusion in clean many-body problems is not natural, although perhaps not forbidden. For the Ising regime of the XXZ model, however, a finite lower bound on the diffusion constant has been found recently\citep{PhysRevLett.119.080602}, thus ruling out subdiffusive spreading. We find the correct picture may involve another peak forming on much lower frequency scales, whose properties cannot be gleaned within the simplistic Ansatz employed thus far --- we return to focus on this issue in Section~\ref{sec:Isingatomic}. 

At the Heisenberg point ($\Delta = 1$) Eq.~\eqref{eq: dynamicalexponent_II} predicts anomalous conductivity with the exponent $\alpha = -3/7$, corresponding to superdiffusive spread of initial distortions according to
%Thus, within our approximation, any magnetic distortion scales with time as 
\begin{equation}
\label{eq: dynamicalexponent_III}
\langle (\Delta x)^2 \rangle \sim t^{10/7} \approx t^{1.43}.
\end{equation}
%\noindent 
%in the long-time limit. 

\paragraph{Comparison with prior results and numerics.% on $\sigma(\omega)$ of XXZ chains
}
%To the best of our knowledge, there are not yet definitive theoretical results let alone consensus on the nature of low finite frequency relaxation in XXZ chains.  
Superdiffusion at the isotropic point has been obtained previously, both analytically with the scaling $\sim t^{12/7} \approx t^{1.71}$ in the continuum 
 limit\citep{PhysRevB.49.3592}, and numerically, by means of the recursion method\citep{PhysRevB.49.15669} ($\sim t^{1.26}$) and  exact diagonalization\citep{PhysRevB.57.8340} ($\sim t^{1.41}$). 
Most recently, $t$-DMRG studies \cite{MarkoAnisotropic, PhysRevLett.117.040601, Ljubotina}%, both in isolated and open system simulations, 
also found superdiffusive transport with $\langle (\Delta x)^2 \rangle \sim t^{4/3} \approx t^{1.33}$. 
Our prediction seems to fall in line with these past numerical results. 
However there are claims, both old and recent, that the spin transport instead has a nondivergent diffusion constant at the isotropic point \cite{deGennes, Mori, PhysRev.138.A608, Auerbach}.

\begin{figure*}[ttp!]
\centering 
\includegraphics[width=18.5cm]{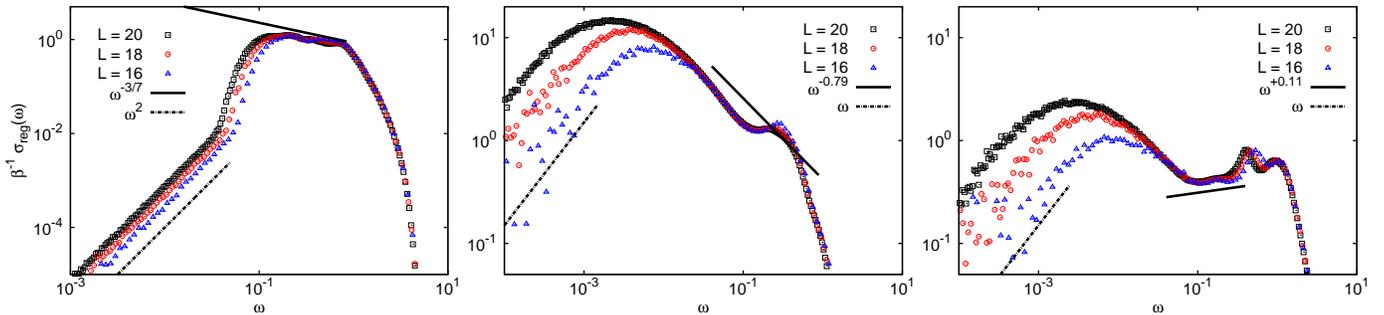}
\caption{Low-frequency Kubo (normalized) conductivity for spin rings of different length, pierced by an irrational magnetic flux and {\it averaged over all magnetization sectors}. The parameters are $\Delta = 1$ and $J_2 = 0.0$ (left), $0.2$ (center) and $-0.2$ (right). The black full lines in the plots correspond to the power laws obtained with the sum-rule method, Eq.~\eqref{eq: dynamicalexponent}, whereas the dotted lines are proportional to either $\omega^2$ or $\omega$. The latter case signals the onset of level repulsion.}
\label{fig:anomalyJ2}
\end{figure*}

We also find good (qualitative and quantitative) agreement with our own numeric studies, as shown in Figs.~\ref{fig: DiffusionExponent}, \ref{fig:anomalyJ2} and \ref{fig: Conductivity_3}. 
See also Appendix~\ref{app: FSEfluxing} where we explore the significance of averaging over magnetization sectors and threading with a flux to lift unwanted symmetries. 
There we also examine the small $\Delta$ regime.

\subsection{Perturbed Heisenberg chains}
\label{subsec:J2}

As already emphasized, it is useful and important to start by visually assessing the overall validity of the three-parameter Ansatz, Eq.~\eqref{eq: ansatz} --- Fig.~\ref{fig: DiffusionExponent} illustrates its predictions at $J_1=\Delta=1$ and $J_2=0,\pm 0.2$, as compared to exact-diagonalization (ED) computations.
The first panel illustrates a surprising sensitivity of the Ansatz parameters on the sign of the next-nearest neighbor spin-flip: positive $J_2$ values enhance superdiffusion, whereas already for relatively small negative $J_2$ long-time transport is suppressed and becomes subdiffusive.
These predictions are borne out in actual numerical results, which can be compared to full Ansatz line shapes (see e.g. right and center panels of Fig.~\ref{fig: DiffusionExponent}, where the ED computation is averaged over the $m_z = 0$ sector).

In order to alleviate finite-size effects we thread the system with an irrational flux (implemented via a Peierls substitution) and work within the grand-canonical ensemble. 
Each of the panels in Fig.~\ref{fig:anomalyJ2} examine both the size and frequency dependence of the Kubo conductivity at $J_2=0, 0.2$ and $ -0.2$, from left to right. 
There are two apparent power laws in each panel: 
(i) the asymptotic finite-size power law which is either $\omega^2$ or $|\omega|$ for integrable or non-integrable models, respectively; the latter being a manifestation of random-matrix like spectral correlations 
(see Appendix~\ref{app: FSEfluxing} for a detailed discussion); 
(ii) mid-frequency power law that persists in the thermodynamic limit, whose dynamic range presumably extends to zero frequency at $J_2=0$, but otherwise is limited by some finite $\omega^*(J_2\to 0)\to 0$, which appears to be masked by finite size effects in our simulations anyway. 
Quite remarkably, the analytically predicted values for the thermodynamic power law, displayed using solid lines, appear to match rather well both the direction and the magnitude of the drift (see also Fig.~\ref{fig: Conductivity_3}). 
Finally, we notice that in the subdiffusive case ($J_2=-0.2$) a new feature appears at very low frequency in the numerics, consistent with a secondary peak forming at much smaller frequencies. This low frequency feature might be an analog of the low-frequency peak we find deep in the Ising regime (see section~\ref{sec:Isingatomic}). It might also be due to strong finite-size effects as we elaborate in Appendix~\ref{app: FSEfluxing}.

\textit{Caveats}. Despite the apparent success of our Ansatz, we do not have any reason to doubt that $\omega^*$, which demarcates the crossover to ordinary diffusion, remains finite in the thermodynamic limit at finite $J_2$, although we do not know how it depends on $J_2$.
%
%In Appendix~\ref{app: FSEfluxing} we further investigate the predictions of the Ansatz, Eq.~\eqref{eq: ansatz}, as compared to ED for different ensembles and symmetries, as well as for other values of $J_2$.

We note that similar anomalous transport properties at low frequencies in \textit{thermal} transport have also been observed in perturbed Heisenberg chains with isotropic NNN coupling\citep{PhysRevLett.96.067202}. 
The anomalies were there shown to be inherited from the integrable point and associated with near-conserved quantities. 
However, in that study the thermal anomaly showed up as a $J_2^{-4}$ effect, which is manifestly independent of the sign of $J_2$. 
Similarly, a $J_2^{-2} - $dependent bound on spin dc conductivity was obtained in a later study using non-local conserved quantities in nonintegrable systems\cite{PhysRevB.76.245108}. 

\section{Two-component response in the Ising (atomic) limit, $\Delta\gg 1$}
\label{sec:Isingatomic}
In the large $\Delta$ limit (for simplicity we only consider $J_2=t_2=0$ --- it is not clear whether integrability breaking is important in this regime) the eigenstates of the model are close to classical spin configurations, i.e.~product states of up-down fully magnetized spins. Due to the locality of the current operator and the product state nature of the $J_1/\Delta=0$ starting point, the conductivity is dominated by local spin flips between, say, sites 2 and 3.  We need to average these processes over \enquote{background} configurations, i.e.~spin configurations on sites 1 and 4 of an infinite chain. The four possibilities are: (i) $... \downarrow \uparrow \downarrow \downarrow ...$, (ii) $... \uparrow \uparrow \downarrow \uparrow ...$, (iii) $... \uparrow \uparrow \downarrow \downarrow ...$ and (iv) $... \downarrow  \uparrow \downarrow \uparrow ...$ %with energies $\ldots$, respectively.
Spin flips between the middle two sites either change the configuration energy by a factor of $\Delta$, or do not change it at all.
To leading order in $J_1$, we ignore energy corrections and simply compute the conductivity in the $\beta\to 0$ limit as
\begin{equation}
\beta^{-1}\sigma'(\omega)=A_0\delta(\omega)+B_0\sum_\pm \delta(\omega\pm\Delta),
\label{eq:acatomic}
\end{equation}
with $A_0=2B_0=\pi J_1^2/8$. More generally, at small but finite $J_1/\Delta$ we still expect to observe this two-component structure of response, with the $\omega=0$ contribution broadened into a Drude-like diffusive peak (with possible anomalies) and a broad absorption peak near $\omega=\Delta$. From the point of view of coarse phenomenological description, we are interested in the evolution (with $J_1$) of amplitudes, widths and locations of the two peaks in Eq.~\eqref{eq:acatomic}. Assuming the low-frequency peak does not move away to a finite frequency, we therefore need five microscopic inputs to extract the five parameters necessary to describe the line shape, which we model with the following simple Ansatz
\begin{equation}
\beta^{-1} \sigma'(\omega)=A\tilde{\delta}_A(\omega)+\frac{B}{2}\sum_\pm \tilde{\delta}_B(\omega\pm\tilde{\Delta}),
\label{eq:largeD}
\end{equation}
with $\tilde{\delta}_A\equiv (\omega_A \sqrt{\pi})^{-1} \textrm{exp}(-(\omega/\omega_A)^2)$ and similarly for $\tilde{\delta}_B$.
This functional form may need to be refined once additional non-perturbative insights (e.g.~from numerics) into the dynamics of this regime become available. As written, the Ansatz is consistent with the current belief of absence of ballistic behavior within the Ising regime of the XXZ chain.
Using the five known moments for the XXZ chain\cite{Roldan} we obtain, by means of  Eq.~\eqref{eq:momenteq1} and \eqref{eq:largeD} 
\begin{equation}
\begin{array}{rclcrcl}
 A &=&  \cfrac{\pi}{8} \, J_1^2 + \cfrac{3 \pi}{32} \left(\cfrac{J_1^2}{\Delta}\right)^2, & \quad &   \omega_A &=& \displaystyle \sqrt{\frac{3}{2}}J_1, \\ \\
 B &=&  \cfrac{\pi}{8} \, J_1^2 	- \cfrac{3 \pi}{32} \left(\cfrac{J_1^2}{\Delta}\right)^2, & \quad &   \omega_B &=& J_1, \\ \\
 \tilde{\Delta} &=& \Delta - \cfrac{1}{4} \cfrac{J_1^2}{\Delta}. & \quad & & &
\end{array}
\end{equation}

With these results Eq.~\eqref{eq:largeD} predicts the diffusion constant $\mathcal{D} \approx 0.18 J_1$  in the limit $J_1\to 0$ (or $\Delta \to \infty$). 
The limiting behavior of the diffusion constant as $\Delta\to\infty$ has received considerable recent attention with earlier studies
(e.g.~Refs.~\citen{MarkoAnisotropic, Steinigeweg}) coalescing around it vanishing $\mathcal{D}\sim 1/\Delta$ while the more recent state-of-the-art t-DMRG studies (both linear response and quench-based probes) 
giving convergence to a finite value\cite{Karrasch, Ljubotina} $\mathcal{D} \approx 0.4J_1$, which is remarkably close to our result, whose only assumption was that the overall shape 
of the AC response is smoothly connected to the analytic result in the extreme limit.  
Intuitively, the physical origin of both the non-vanishing diffusion constant and even the presence of the low frequency peak is in the finite density of thermally excited voids in which 
magnetization can fluctuate.

\begin{figure}[ttp]
%\centering 
\includegraphics[width=8.5cm]{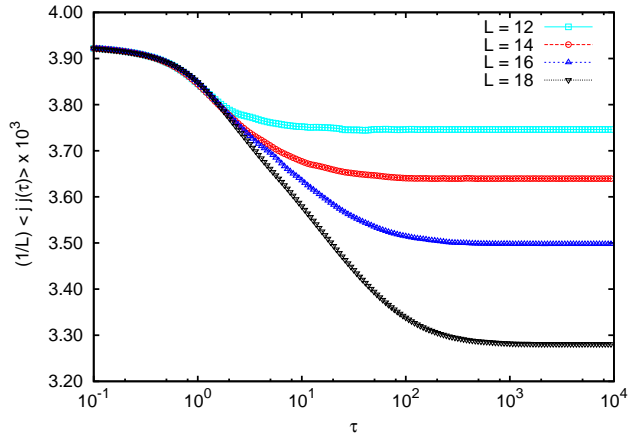}
\caption{Temporal and size dependence of current auto-correlation decay for weakly coupled even-odd sublattices (Large $J_2$ limit) on a semi-log scale. The long-time plateau indicates a finite Drude weight which decreases with increasing system size, whereas the straight line decay (i.e.~$\log{t}$) becomes more pronounced.}
\label{fig:logt}
\end{figure}

%Th saturation of the diffusion constant in the Ising limit (independence with $\Delta$) is in agreement with long-time : both nonequilibrium quenches and 
%invoking of Einstein relations for correlation functions \cite{Karrasch, Ljubotina} predict such a saturation, with .
%This is however inconsistent with earlier studies, e.g. Refs. \citen{MarkoAnisotropic, Steinigeweg}, which predict that the diffusion constant inversely vanishes with $\Delta$ in the Ising limit.
%
%We note, however, that this is in disagreement with Ref. \citen{Karrasch} which suggests the diffusion constant saturates instead. 

\section{Nearly free fermions}
\label{sec: NearlyFree}
The model Hamiltonian $\hat H_0+\delta \hat H_1$ has several free-fermion limits. Here we examine two of them: $\Delta/J_1\to 0$ and $J_1/J_2=\Delta/J_2\to 0$.

{\it $\Delta \ll 1$ limit of the integrable chain} --- We have shown in Sec.~\ref{sec: analytic_results} the Ansatz defined by Eq.~\eqref{eq: ansatz} 
predicts the low-frequency %subleading 
behavior $\sigma'(\omega) \sim D \, \delta(\omega) + |\omega|^{\alpha}$ for the spin conductivity for $\Delta < 1$, 
interpolating monotonically from $\alpha=-3/7$ (Heisenberg chain) to $\alpha=-1$ (XY chain), with $\Delta \to 0$ attained as
%with the dynamical exponent lying within the interval $-1 < \alpha < -3/7$. 
%
%More precisely, the {\it regular component} of the conductivity shows sub-ballistic behavior, smoothly going towards ballistic-like scaling of the exponent $\alpha$ as $\Delta$ approaches zero
\begin{equation}
\label{eq:smallDapprox_II}
%\beta^{-1}\sigma'(\omega) \approx \frac{2 \pi}{5} \frac{\Delta^2}{\omega^{1 - \frac{4\Delta^2}{5}}} e^{-\frac{4}{5}\omega^2} + \mathcal{O}\left(\Delta^4\right), \quad \mbox{for} \quad \Delta \ll 1.
 \beta^{-1} \sigma'(\omega) \approx \frac{2\pi}{5}\frac{\Delta^2}{\omega^{1-\frac{4\Delta^2}{5J_1^2}}}\textrm{exp}(-\frac{4}{5}\omega^2).
\end{equation}
Note that lack of normalizability of the $|\omega|^{-1}$ power law in the limit $\Delta\to 0$ is avoided via a vanishing coefficient at the level of the Ansatz.

{\it Large $J_2$ limit} --- For large $J_2$ (with $J_1=\Delta=1$) the spin chains decouple into two independent chains.
Moreover, from the results shown in the left panel of Fig.~\ref{fig: DiffusionExponent}, we expect the $\pm J_2$ dynamical responses to be identical as $J_2 \rightarrow \infty$, 
with $\alpha \rightarrow -1$, as in the small $\Delta$ limit treated in the preceding paragraphs.
Note nevertheless that for $1 \ll J_2 \neq \infty$ the system is still nonintegrable and therefore normal diffusion is expected at long times. In this limit, and for $\omega \ll \omega_0$ %for $\omega_0 \to \frac{\sqrt{5}}{2}$ and
\begin{equation}
\label{eq: largeJ2}
%\beta^{-1} \sigma(\omega) \approx \frac{49 \pi}{10} \frac{1}{\omega^{1-\frac{49}{20J_2^2}}},
 \beta^{-1} \sigma'(\omega) \approx \pi \frac{\kappa}{\omega^{1 - \frac{\kappa}{2J^2_2}}},
% + \mathcal{O}\left(\frac{1}{J_2}\right), \quad \mbox{for} \quad |J_2| \gg 1;
\end{equation}
with $\kappa =  \frac{25J^4_1 + 20J^2_1\Delta^2 + 4\Delta^4}{7J^2_1 + 3\Delta^2}$. This result has the same structure as that of Eq.~\eqref{eq:smallDapprox_II} when reinterpreted as perturbation theory in $1/J_2$.

Before we present numerical evidence supporting these results, let us remark that $|\omega|^{-1}$ behavior is profoundly different from the Lorentzian line shape we expect from perturbative calculations, 
e.g.~it corresponds to logarithmically slow decay in time! Such slow decay must be tied in to the emergence of some conserved quantity we have yet to identify. 
We also caution that the approach to the non-interacting limit may or may not follow the prediction of the Ansatz in detail. 
For example, there may be a universal (interaction independent) $|\omega|^{-1}$ tail, limited both at low and high frequencies by interaction-dependent cutoffs, instead of interaction-dependent exponent. 
Our numerical tests are too coarse to differentiate this possibility from the varying exponent.  
%
%***i may be too pessimistic perhpas?*******
%
These are instead primarily focused on documenting the decay of the conductivity, which \emph{approximately} behaves as $|\omega|^{-1}$ both near $\Delta=0$ and in the large $J_2$ limit, with concomitant logarithmic decay in temporal current auto-correlations. 
The latter is depicted in Fig.~\ref{fig:logt} and appears rather compelling. This figure also displays the remnant Drude peak in these finite-sized samples, which appears to decay exponentially with system size, consistent with the lack of integrability in that particular limit. 
In Appendix~\ref{app: weaklyinteracting} we present data in the frequency domain which corroborates this picture, albeit with significantly stronger finite size effects for the integrable case. 
\section{Non-linear hydrodynamics in strongly ergodic chains}
\label{sec:nonlinhydro}
Weakly non-integrable, strongly Ising and nearly non-interacting limits (Sections III, IV, V, respectively) all benefit from some analytic insight to guide the Ansatz-based approach used in this work. Away from all such limits we expect current auto-correlations to exhibit the least complicated structure possible for a non-conserved operator, i.e.~monotonic decay with no particular short or long transient time scales. Importantly, such decays need not be simple exponential, e.g. as dictated by conventional Drude line shape and textbook hydrodynamics. Long-time tails can originate from non-linear admixture of \enquote{slow} conserved operators into everything else of interest, including nominally \enquote{fast} currents. Intuitively, this may be seen as follows: suppose we postulate ordinary diffusion with concomitant local Fick law relationships between currents and density gradients, $j$ and $n$, respectively (both in principle with multiple components, e.g.~magnetization, energy, etc.)
\begin{equation}
j(r,t)=-\mathcal{D} \, \nabla n(r,t)+\eta(r,t),
\label{eq:fick}
\end{equation} 
with $\eta$ denoting fast uncorrelated noise and $\mathcal{D}$ the diffusion (matrix) constant. 

More generally, however, $\mathcal{D}$ is itself a symmetry-constrained operator, $\mathcal{D}(r,t)=\mathcal{D}_0+C n(r,t)+\ldots \, $, where $C$ is a certain third rank constant, $n(r,t)$ denotes spatio-temporal deviations of the conserved quantities from equilibrium and \enquote{$\ldots$} denotes other higher powers of conserved quantities which are less relevant near the diffusive fixed point.
Thus $\mathcal{D}$ inherits slow fluctuations from the conserved quantities, thereby inducing power-law corrections to the presumed fast current decay. 
These nominally perturbative fluctuation effects have been known
%https://journals.aps.org/prl/abstract/10.1103/PhysRevLett.25.1257
 since the early 1970's\cite{Dorfman1970} %(and presumably even earlier, in ordinary fluid hydrodynamics) 
but their qualitative and quantitative significance (especially in low-dimensional and/or disordered media) is a subject of ongoing research. 
Most relevantly, Mukerjee et.al. \cite{PhysRevB.73.035113} observed that non-integrable fermionic chains of the type defined above attain a strongly non-linear hydrodynamic regime, whereby the long-time tail 
\begin{equation}
\lim_{\beta \to 0}\beta^{-1}(\sigma'(\omega)-\sigma'(0)) \sim -\sqrt{\omega},
\label{eq:}
\end{equation} 
completely dominates over most of the dynamical range. Thus, to a good accuracy
\begin{equation}
\langle \hat j^z(t) \hat j^z(0)\rangle=\frac{a}{(\tau^2+t^2)^{3/4}},
\label{eq:2parameter}
\end{equation} 
with $a$ and $\tau$ both of order one in natural units. We note in passing that this functional form also implies that the extreme high-frequency tail must be exponential ---  this is indeed the case but we defer discussing it to a later date. 

\begin{figure}[ttp!]
\centering 
\includegraphics[width=8.5cm]{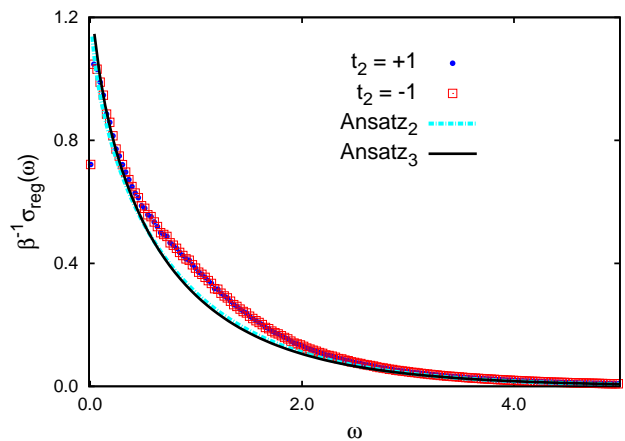}
\caption{
%2parameter (green): $1.62 \omega^{0.25} K_{-0.25} (0.87 \omega)$, 3parameter form (black): $3.93 e^{-0.92 \sqrt{\omega} - 0.65 {\omega}}$
Kubo conductivity for a spinless fermion model on a 18-site chain with next-nearest neighbor hopping amplitudes $t_2= \pm 1$ and nearest-neighbor interaction $V=2$
(uniform binning, data normalized to one). The three lines are from 2 and 3 parameter ans\"{a}tze (described in text).
}
\label{fig:fermionic}
\end{figure}

We now proceed to provide a more microscopic %detailed and quantitative 
account of the conductivity in this regime. We start by displaying the first three moments of the conductivity at $J_2=0$
\begin{eqnarray}
\label{eq: mfermion1}
\mu_1 & =  &  2\, (t_1^2+4 \, t_2^2), \\ \nonumber \\
\label{eq: mfermion2}
\mu_2 & =  &V^2 ( t_1^2 + 4 \, t_2^2), \\ \nonumber \\
\label{eq: mfermion3}
\mu_3 & = & V^2 (5 t_1^4 + 24 t_2^4 + V^2 ( t_1^2 + 4 t_2^2) + 30 t_1^2 t_2^2), 
%\raggedleft
\end{eqnarray}
where we used the more familiar notation for the fermionic model $t_1=J_1$ and $V=2\Delta$.

The simple 2-parameter Ansatz in Eq.~\eqref{eq:2parameter} may be Fourier transformed as 

\begin{equation}
\beta^{-1}\sigma_1'(\omega) = a (\tau \omega)^{1/4} K_{-1/4}(\tau\omega).
\end{equation}
This expression is overly simplistic, however, as it fixes both the high-frequency envelope and the amplitude of the low-frequency non-analyticity with a single constant.
%, which may be straightforwardly checked against exact microscopic moments. 

Another more general Ansatz, and one that separates low- and high-frequency behaviors explicitly, may be devised, e.g. 
 \begin{equation}
 \label{eq:fermion3ansatz}
  \beta^{-1} \sigma_2' (\omega)= A \exp(-b \sqrt{\omega}-c \, \omega).
 \end{equation}

We can use the first two and all three of the moments in Eqs.~(\ref{eq: mfermion1} -- \ref{eq: mfermion3}), respectively, to numerically extract the parameters in both Ans\"{a}tze and plot them together with the exact (finite-size) conductivity.
The 2- and 3-parameter Ans\"atze are shown in Fig.~\ref{fig:fermionic}.
It is rather clear that both are very close --- the exact conductivity is subjected to finite-size effects, most pronounced as the pseudogap at low frequency, but also present elsewhere\cite{PhysRevB.73.035113}.

We conclude with some comments about the weak integrability-breaking regime of this model, i.e.~$t_2\to 0$.  First, the presence of the additional $\hat S^z$ operator in the perturbation carries an important constraint, namely it implies that symmetric averages over positive and negative magnetization sectors will be symmetric under $t_2\to - t_2$ transformations. This may be seen very clearly already in the short-time expansions, where terms odd in $t_2$ are absent (unlike the $J_2$ case. See Eqs.~(\ref{eq: mfermion1} -- \ref{eq: mfermion3})). This alone does not preclude, of course, some version of the phenomena observed in Section III, e.g.~drifting exponents, which may be predicted following the same steps as before. Our attempt to observe shifting power laws using exact diagonalization (as we did in Section III) failed --- as best as we can tell the anomaly of the Heisenberg point gives way to hydrodynamic behavior at already relatively weak $t_2$, with no discernible change in the exponent over intermediate range of frequencies.

\section{Implications for experiments -- the structure factor}
\label{sec:experiments}
We have presented several examples of low-frequency conductivity anomalies, which may in principle be observed in quasi one-dimensional materials, spin-chain compounds or cold atomic gases. The probing frequency should not be too low, as there are numerous residual weak couplings to the environment, e.g. phonons and disorder, which should in principle modify the behavior at too low frequencies. Quite generally, it is difficult to excite a many-body system in a controlled linear response regime. It is often easier to observe slow fluctuations, e.g. using inelastic light or neutron scattering, time-of-flight techniques, etc. 
The relevant quantity is the so called dynamic structure factor, which may be related (as shown in Appendix~\ref{app: apendix_memoryfunction}) in the low-frequency long-wavelength limit to the Kubo conductivity as 
\begin{equation}
S(q,\omega) = \frac{ q^2 \sigma'(\omega)}{ \left( \omega + q^2 \displaystyle \frac{\beta}{\chi_0} \sigma''(\omega) \right)^2+\left(\displaystyle\frac{q^2}{2} \frac{\beta}{\chi_0} \sigma'(\omega) \right)^2},
\label{eq:DSF}
\end{equation}

\begin{figure}[ttp!]
\centering 
\includegraphics[width=8.5cm]{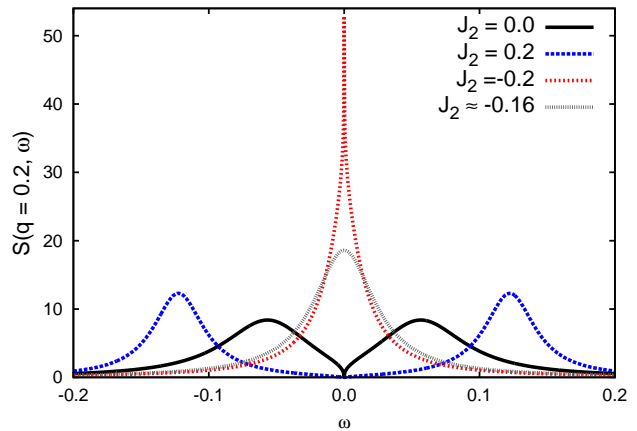}
\caption{Dynamic structure factor for %frequencies $\omega < 
$q = 0.2$ in chains with $\Delta =1 $ and four values of $J_2$. Note the distinctive non-Lorentzian line shapes,
%spectral signatures between 
quasi-Bragg in the the sub-diffusive case ($J_2=-0.2,\ \alpha\approx 0.11$) and double-peaked in the super-diffusive cases (with exponents $\alpha=-3/7$ and $\alpha\approx -0.79$, for $J_2=0, 0.2$, respectively), as compared to the normal diffusive case ($J_2 \approx -0.162$).}
\label{fig:structure_factor}
\end{figure}

\noindent 
with $\chi_0$ the static spin susceptibility and $\sigma''(\omega)$ the imaginary part of the Kubo conductivity, related to $\sigma'(\omega)$ through the Kramers-Kronig relation $$\sigma''(\omega) =\mbox{P} \displaystyle \int \frac{d \omega'}{2\pi} \frac{\sigma'(\omega')}{\omega'-\omega},$$
where $\mbox{P}$ indicates the Cauchy principal value. 
%Eq. \eqref{eq:DSF} is the finite-$q$ generalisation of Eq. \eqref{eq:eom}. 
%The $\sigma''(\omega)$ piece may be neglected in the small $(q, \omega)$ regime for subdiffusive and diffusive transport, leading to the usual ``diffusion'' pole formula, and provides a good approximation to $S(q,\omega)$. For superdiffusive transport, its contribution is significant. 
%We observe that t
The non-diffusive low-frequency anomalies discussed above, $\sigma'(\omega)\sim |\omega|^\alpha$, which are accompanied by 
$$\sigma''(\omega)\sim \mbox{sgn}(\omega) |\omega|^\alpha \tan( |\alpha| \frac{\pi}{2})+\omega \left(\frac{\alpha-3}{4}\right) \Gamma\left(\frac{\alpha-3}{2}\right), $$ 
\noindent
lead to very distinctive spectral signatures in $S(q,\omega)$: 

\textit{a.  Sub-diffusion} ($1>\alpha>0$) results in a quasi-elastic Bragg-like peak, with a doubly-divergent form
\begin{equation}
\lim_{\omega \to 0}S(q,\omega)\sim \frac{1}{q^2 |\omega|^{\alpha}};
\end{equation}

\textit{b. Super-diffusion} ($-1<\alpha<0$) splits the central peak into a doublet, which also disperse albeit non-ballistically, with new peak locations at 
\begin{equation}
\omega_\pm (q) \approx \pm \left(\frac{\, q^2 \tan ( |\alpha| \frac{\pi}{2})}{2+|\alpha|} \right)^{\frac{1}{1+|\alpha|}}\sim\pm |q|^{\frac{2}{1+|\alpha|}}.
\end{equation}

The structure factor in this case vanishes at the origin as $S(q,\omega\to 0) \sim q^{-2} |\omega|^{|\alpha|}$ and diverges at the peaks $S(q\to 0, \omega_\pm)\propto \omega_\pm^{-1}$. Unlike coherent peaks that exhibit asymptotic narrowing and divergent quality factors (e.g.~as arising from Goldstone modes), here the quality factor remains finite, except at $\alpha\to -1$.

We expect these qualitative differences to be helpful in identifying anomalous behavior in future experimental and numerical studies. 

Let us close this section with two further comments: (i) the analysis of $S(q,\omega)$ above heavily relies on the small momentum ($q\to 0$) expansion of the memory function in the denominator. It is in principle possible that the analyticity in $q$ is compromised by the presence of non-diffusive behavior, e.g.~if higher order terms in $q$ change the asymptotic behavior at small frequencies. We have partially investigated and addressed this concern by explicitly computing the next term in the small-$q$ expansion (following the same procedure as for the $q=0$ conductivity and using Eq.~\eqref{eq: nestedSz}) and estimating its anomaly structure.  We found that the anomaly is weaker than that of the leading term, so the leading term (in the small momentum expansion) continues to dominate at low frequencies, at this level of approximation; (ii) in addition to the \enquote{central} Rayleigh peak ($\omega\ll q$), conventional fluids also exhibit "phonon" (Brillouin) peaks at frequencies $\pm c \, q$ (with $c$ the speed of sound) whose width vanishes as $q^2$. It is worth recalling that the presence of sound waves in conventional fluids necessarily comes with a $\delta(\omega)$ in the conductivity --- both peaks are mandated by the exact momentum conservation in free space. This picture lends itself to the following generalization which we believe takes place in integrable lattice models, where the conductivity exhibits both regular and Drude components: at finite momenta the coherent zero-frequency (Drude) weight disperses rapidly to a relatively high frequency $\sim q$, whereas the regular part associated with (super-) diffusion is spread out to a much lesser extend, e.g.~$q^2$ for ordinary diffusion. 
The presence of dispersing quasi-Brillouin peaks is consistent with earlier numerical studies \cite{PhysRevB.77.245131} and also exact results on the XY model\cite{Katsura, McCoySpin}.
Intriguingly, the additional superdiffusive peaks we find above disperse parametrically slower than $q$ and therefore can in principle coexist with quasi-phonon peaks in the structure factor. Clearly, the dynamical structure factor can display a much richer behavior than $q=0$ conductivity itself and we plan to return to this subject in the near future\cite{dispersion}.

\section{Discussion and outlook}
\label{sec:conclusion}
This work has assembled a set of diverse results on dynamical signatures of transport anomalies in integrable, weakly and strongly non-integrable spin chains. The significance of these results, many of which are new and some are surprising, is amplified by the fact that they appear in relatively common quantum models at finite (high) temperature, and may be directly observed in transport or scattering probes.  Particularly intriguing are the continuously drifting exponents (Section III) in nearly integrable cases, and the logarithmic in time relaxation (Section V) in weakly interacting chains (and ladders).  We have also included a detailed discussion of implications of these result to the dynamical structure factor, including likely existence of the additional phonon-like Brillouin peak in integrable cases, in Section VII.

There are several open questions and directions of inquiry worth mentioning. To start, it is unclear whether the presence of additional conservation laws leads to a hierarchy of dynamic anomalies -- certainly in the specific case of integrable XXZ chains the exact conservation of the energy current implies that the response of that particular channel remains purely ballistic (no finite frequency absorption whatsoever). One potentially natural way to approach these questions is from the vantage point of 
generalized hydrodynamics, in terms of conserved quantities and their effective velocities, as recently formulated\cite{Doyon, PhysRevLett.117.207201}. However, it is at present unclear whether that formalism is capable of addressing the regimes of interests to us, e.g. linear response near infinite temperature and for the entire range of $\Delta$. 
Recent results\cite{PhysRevLett.119.020602, PhysRevB.96.115124, PhysRevB.97.045407} appear encouraging, although restricted to \textit{nonequilibrium transport}. 
In the meantime it is also feasible to apply the same methodology as in this work (e.g.~moments+Ansatz+exact diagonalization) to other response functions, e.g.~heat conductivity and thermoelectric response, also finite momentum response (including the dynamic structure factor) \emph{and} as a function of temperature, to provide further data for the more macroscopic hydrodynamic theory. 
%We aim to pursue both approaches in the near future.  
It would be interesting, in particular, to explore the role of symmetries in determining the anomaly structure. For example, empirically (and somewhat surprisingly), we found no difference between perturbing the Heisenberg point with nnn XY coupling vs Heisenberg.

In separate follow-up projects we will present (i) detailed studies of high frequency tails of response functions, both in integrable and strongly non-integrable chains, using resummation techniques. Our current conjecture is while the latter is generically simply exponential, integrable chains have sufficiently sparse matrix elements to allow for functionally distinct, e.g. Gaussian, fall-off of the response; (ii) linear response anomalies should also be accompanied by non-equilibrium signatures, e.g. in current noise as probed by non-equilibrium steady states' fluctuations.

Lastly, our early motivation for this project came via many-body localization in disordered chains (in turn motivated by similar $J_2-$sign effects on transport at zero temperature discovered earlier \cite{PhysRevA.92.013618}). 
Localization phenomena had to be set aside as we kept unearthing rich structures in the clean problem. We plan to re-engage many-body localization using methods of this paper.

We thank A. Abanov, A. Auerbach, S. Gopalakrishnan, D. Huse, K. Jensen, R. Konik, H. Monien, D. Pekker, T. Prosen, M. Rigol, A. Rosch, S. Sondhi, A. Tsvelik, and M. \v Znidari\v c for discussions.
VKV thanks ICTP, Trieste for hospitality and computing resources where this work was initiated. 
We gratefully acknowledge support from the Simons Center for Geometry and Physics, Stony Brook University at which some of the research for this paper was performed. 
VKV and VO also acknowledge support from the NSF DMR Grant No. 1508538 and US-Israel BSF Grant No. 2014265.
RJS acknowledges support from the H2 branch of the Bonn-Cologne Graduate School of Physics and Astronomy. 
We benefited from benchmarking some of our results against the DiracQ package~\cite{DiracQ}.

\appendix 

\section{Construction of the memory function}

In this appendix we derive the so-called memory function\citep{Mori} of the spin autocorrelation which will allow us to (i) relate the spin spectral function with the Kubo conductivity through the exact expression~\eqref{eq:DSF} [see Eq.~\eqref{eq: Laplace_EOM}]; and 
 to (ii) re-derive the generator of the sum-rule relation~\eqref{eq:eom} [see Eq.~\eqref{eq: sum-rules-2}].
We follow the notation and procedure of Berne and Harp\citep{Berne}. 

To construct the memory function consider the set $\{\hat Q_j \}$ of operators describing the slow modes of our model, with $\hat Q_0 \equiv \hat S_q^z$ the Fourier transform of  
$\hat S^z_i$ and $\hat Q_{j \, >0}$ the (infinite) set of local (or quasilocal) conserved quantities associated with the integrable limit of the Hamiltonian, Eq.~\eqref{eq: hamiltonian}. 
Note that we do not need to include in $\{\hat Q_j \}$ the operator describing the slow energy fluctuations of the system since these do not couple to the magnetization, 
and hence the set $\{\hat Q_j \}$ only includes $\hat Q_0$ in the nonintegrable case. We choose the $\hat Q_{j \, >0}$ to be orthogonal to each other with respect to the scalar product in operator space

\begin{equation}
(\hat Q_i | \hat Q_j ) = Z^{-1} \mbox{Tr} (\hat Q_i \hat Q_j) = \delta_{i, j} \, \chi_{i} \, \beta^{-1},
\end{equation}

\noindent 
with $\chi_{j} = \beta (\hat Q_j | \hat Q_j)$ the static susceptibility associated to $\hat Q_j$ and $Z=\mbox{Tr} e^{-\beta \hat H}$ the partition function of the model. Furthermore the $\{\hat Q_j \}$ are so chosen that their ensemble average are zero, 
that is $\langle Q_j \rangle = Z^{-1} \mbox{Tr} \hat Q_j  = 0$. We can then regard the set $\{\hat Q_j \}$ as a set of vectors in a Hilbert space and  define the projector operator $\hat P$ onto the subspace spanned by $\{\hat Q_j\}$ as 

\begin{equation}
\hat P = \sum_j \beta \chi_j^{-1} | \hat Q_j )( \hat Q_j |.
\end{equation}

To study spin dynamics we consider the Heisenberg equation of motion $\partial_t | \hat Q_0 (t) ) = i \hat L | \hat Q_0 (t) )$, where $\hat L$ is the Liouville operator defined by $i \hat L \hat O \equiv -i [\hat O, \hat H]$. This equation can be rewritten as 

\begin{equation}
\partial_t | \hat Q_0(t) ) = i \hat L \hat P | \hat Q_0(t) ) + i \hat L (1- \hat P) | \hat Q_0(t) ),
\label{eq: EOM_vector}
\end{equation}

\noindent 
and so, the equation of motion describing the spin autocorrelation function $S(q, t) \equiv (\hat Q_0|\hat Q_0(t))$ follows

\begin{equation}
\begin{array}{rcl}
\partial_t S(q, t)  & = &     (\hat Q_0 | i \hat L \hat P | \hat Q_0(t) ) \\ \\
                           &    &  + \, (\hat Q_0 | i \hat L (1- \hat P) | \hat Q_0(t) ).
\end{array}
\label{eq: EOM_pre}
\end{equation}

The first term on the right hand side of Eq.~\eqref{eq: EOM_pre} vanishes since $i L | \hat Q_j ) =  \delta_{0, j} | i L \hat Q_j )$, given that the $Q_{j \, > 0}$ are constants of motion and $(\hat Q_0 | i L| \hat Q_0) = 0$, which follows because $Q_0$ is odd under time reversal. On the other hand, an expression for the second term on the right hand side of Eq.~\eqref{eq: EOM_pre} can be obtained by acting with the operator $(1- \hat P)$ on the left of Eq.~\eqref{eq: EOM_vector} and solving for $(1- \hat P) | \hat Q_0(t) )$. This yields

\begin{eqnarray}
(1- \hat P) | \hat Q_0(t) )  & = & \sum\limits_j \int_0^t d\tau e^{i(1-\hat P) \hat L (t-\tau)}(1-\hat P) i \hat L |\hat Q_j ) \nonumber \\
 & & \times \beta \chi_j^{-1}(\hat Q_j | \hat Q_0 (\tau) ).
\end{eqnarray}

Noting that only the $j=0$ contribution from the summation above is finite, and that

\begin{equation}
e^{i (1-\hat P) \hat L t}(1-\hat P) = (1-\hat P) e^{i (1-\hat P) \hat L (1-\hat P) t},
\end{equation}

\noindent
we write Eq.~\eqref{eq: EOM_pre} as

\begin{eqnarray}
\partial_t S(q, t)  & = & -\int_0^t d\tau (\hat Q_0| i \hat L (1-\hat P) e^{i(1-\hat P) \hat L (1-\hat P)(t-\tau)} \nonumber \\  \nonumber \\ 
 & & \times (1-\hat P) i \hat L |\hat Q_0 ) \beta \chi_0^{-1} S(q, t).
 \label{eq: EOM_pre2}
\end{eqnarray}

Finally, using the continuity equation

\begin{equation}
 i\hat L \hat Q_0  = -i q \, \hat j_q,
\end{equation}

\noindent
where $\hat j_q$ is the Fourier transform of the total spin current, one can rewrite Eq.~\eqref{eq: EOM_pre2} as 
\begin{equation}
\partial_t S(q,t) + q^2 \int_0^t d \tau \Sigma(q, t-\tau) S(q,\tau) = 0,
\label{eq: EOM}
\end{equation}
with $\Sigma(q, t)$ the memory function\citep{Mori}, defined here as 

\begin{eqnarray}
\beta^{-1} \chi_0 \, \Sigma(q, t)  = ( \hat j_q | (1-\hat P) e^{i (1-\hat P) \hat L (1-\hat P) t} (1-\hat P) |  \hat j_q), \nonumber \\ \nonumber \\
 &    & 
 \label{eq: memoryfun1}
\end{eqnarray}

From Eq.~\eqref{eq: memoryfun1} we see the memory function differs from the usual current-current correlation function in two aspects\citep{Forster}: first, the evolution operator determining the spectrum of $\Sigma(q,t)$ has the intrinsic fluctuations of the slow modes $\{\hat Q_i \}$ projected out of it; second, only the components of the current orthogonal to the subspace spanned by the $\{\hat Q_{i>0} \}$ determine the memory function. 

In the long-wavelength limit, however, one can show the projection operation has no effect\citep{Forster} and so

\begin{eqnarray}
\lim\limits_{q \rightarrow 0} \beta^{-1} \chi_0 \, \Sigma(q, t) & = & \lim\limits_{q \rightarrow 0} \, \, ( \hat j_q | e^{i \hat L t} |  \hat j_q) \nonumber \\ \nonumber \\ 
 & = & \sigma(t),
 \label{eq: hydrodynamic-limit}
 \end{eqnarray}
 
\noindent 
with $\beta^{-1} \chi_0 = (\hat S_q^z| \hat S_q^z) = 1/4$ and $\sigma(t)$ the Kubo conductivity. 
%*** VO: I believe $\sigma=\beta \int_0^\infty dt \langle j(t) j(0)\rangle$, with finite frEq.~related through cosine FT transform for real part of $\sigma(\omega)$. if that's correct and $C(k,0)\equiv C=1/4$ (see below Eq.~8) then we know how to fix Eq.~6**** 

Eq.~\eqref{eq: hydrodynamic-limit} implies the memory function displays the same behavior as the Kubo conductivity in the long-wavelength limit. That the memory function is well behaved as $q$ approaches zero follows since the total spin current operator is a local operator and, consequently, its correlation function falls off rapidly with distance. A bit more subtle is the long-time limit, at which the memory function might become nonanalytic due to, e.g., (i) the aforementioned presence of local or quasilocal conserved quantities coupling to the current, which contribute to $\Sigma(q,t)$ in the long-wavelength limit, or (ii) mode-mode coupling of the conserved spin density, which may be relevant even at finite wavevectors\citep{Forster}. 

To derive Eq.~\eqref{eq:DSF} in the main text, consider the Laplace-transform of Eq.~\eqref{eq: EOM}

\begin{equation}
\tilde S(q,z) = \frac{i \beta^{-1}}{z+iq^2 \tilde \Sigma(q,z)} \chi_0, \quad (\mbox{Im} z > 0),
\label{eq: Laplace_EOM}
\end{equation}

\noindent
and note that in the long-wavelength limit $\tilde S(q,z)$ has a pole at $z=0$, which is a direct consequence of the fact that $\lim\limits_{q\rightarrow 0} S(q, t)$ does not decay in time. 

The spectral function is then given by

\begin{equation}
S(q, \omega) = \lim_{\epsilon \to 0} \left[ \tilde S(q, \omega+ i \epsilon)- \tilde S(q, \omega- i \epsilon) \right],
\end{equation}

\noindent
from which Eq.~\eqref{eq:DSF} directly follows, after using the limiting expression Eq.~\eqref{eq: hydrodynamic-limit}.

Now, using the representation 

\begin{equation}
2\pi i \tilde f(q,z) = \int d\omega f(q, \omega)(\omega-z)^{-1},
\end{equation}

for both $\tilde S(q, z)$ and $\tilde \Sigma(q, z)$ one can, in the limit of large $z$, Taylor expand both sides of the dispersion relation, Eq.~\eqref{eq: Laplace_EOM}, in $z^{-1}$ and equate the expansion coefficients to obtain a set of identities relating the frequency moments of the spin correlation function with those of the memory function. We write these sum rules as

\begin{eqnarray}
\raggedleft
\mu_n & \equiv & \frac{1}{q^2}\int \frac{d\omega}{2\pi} \omega^{2n} S(q, \omega) \nonumber \\ \nonumber \\ 
 & = & \int \frac{d\omega}{2\pi} \omega^{2n-2} \beta^{-1} \chi_0\Sigma(q, \omega),
\label{eq: sum-rules-2}
\end{eqnarray}

\noindent
which are, of course, equivalent to Eq.~\eqref{eq:eom}.

\section{Spectral transfer in a trimer}
\label{app: trimer}
Our Ansatz-based results for the $J_2$-dependence of the conductivity attain a particularly simple form in the few-body limit, e.g.~a trimer. Consider a 3-site spin chain, with two up spins and one down spin. This system may be readily shown to respond, through the current operator, Eq.~\eqref{eq:current}, only at two frequencies 
\begin{equation}
 \label{eq: trimer}
 \omega_{\pm} = \frac{1}{8}\left ( \Delta - 6J_2 \pm \sqrt{(\Delta + 2J_2)^2 + 32} \right ),
\end{equation}

\noindent
with the conductivity given by

\begin{widetext}
\begin{equation}
\scalebox{0.94}{
$
\beta^{-1} \sigma'(\omega)  =  \cfrac{\omega^2_+ \delta(\omega - \omega_+)}{32 + 4(2J_2 + \Delta)^2 + 4(2J_2 + V)\sqrt{(2J_2 + \Delta)^2 + 32}}  + \cfrac{\omega^2_- \delta(\omega - \omega_-)}{32 + 4(2J_2 + \Delta)^2 - 4(2J_2 + \Delta)\sqrt{(2J_2 + \Delta)^2 + 32}}, 
 $
 }
\label{eq: trimer2}
\end{equation}
\end{widetext}

%\vspace{0.2cm}

\noindent
for small $J_2 \ll \Delta$ and $\omega_- < \omega_+$ --- with this difference increasing for $J_2 < 0$. 
Thus for positive $J_2$ the two frequency modes are constrained to be closer to each other, and with about the same spectral weight, whereas for negative $J_2$ the two modes separate out into a high- and a low-frequency mode, the latter being spectrally suppressed in amplitude.

This means that in the presence of small negative next-nearest neighbor spin flips the system's dynamics gets slower due to the spectral redistribution. Such redistribution is predicted by our memory function Ansatz and is vindicated by the numerical linear response calculations, as shown in Fig.~\ref{fig: DiffusionExponent}.

Note that setting $\Delta=0$ in the above Eq.~\eqref{eq: trimer} and \eqref{eq: trimer2} gives identical dynamical properties for both signs of $J_2$. This can also be seen directly in the first three moments,  Eq.~(\ref{eq:m1} -- \ref{eq:m3}): the odd powers of $J_2$ are always coupled to $\Delta$. Hence, it is really the competition between $J_2$ and the anisotropy that begets the anomalous dynamics.
A similar competition in the system's ground state also gave rise to increased ballistic transport in the frustrated ($J_2>0$) case as compared to the unfrustrated case\cite{PhysRevA.92.013618}.

\section{Approaching free-particle limits}
\label{app: weaklyinteracting}
\label{app: BallisticTransport}

\label{app: apendix_memoryfunction}
\begin{figure*}[tthp!]
\centering 
\includegraphics[width=16.5cm]{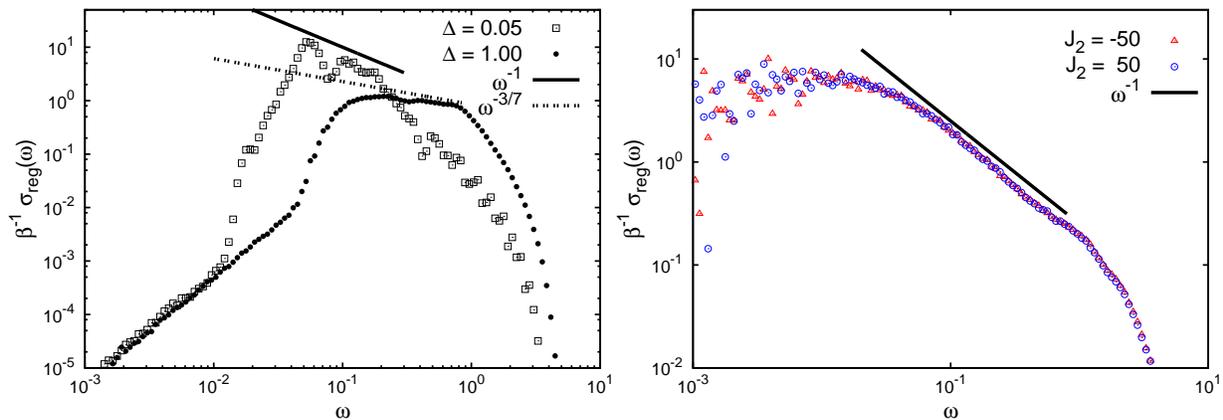}
\caption{$L=18$ Kubo (normalized) conductivity of fluxed chains in the grand canonical ensemble for extreme limits of Hamiltonian parameters. 
\textit{Left panel}: the integrable chain comparing the small $\Delta$ limit with the isotropic point, with the former fitting a steeper power-law ($\approx \omega^{-1}$) at mid-frequencies as predicted by the 
Ansatz Eq.~\eqref{eq:smallDapprox_II} before a $\Delta^2$ drop (which occurs well before the onset of $\omega^2$ finite-size effect as in the isotropic point) sets in, which is also in turn in line with the ansatz 
prediction Eq.~\eqref{eq:smallDapprox_II}.
\textit{Right panel}: the nonintegrable chain for large $J_2$ limit where the ansatz (Eq.~\eqref{eq: dynamicalexponent} and left panel of Fig. \ref{fig: DiffusionExponent}) predicts an insensitivity to sign of 
$J_2$. This latter prediction and the associated power-law $\approx \omega^{-1}$ given by Eq.~\eqref{eq: largeJ2} is vindicated by the displayed numerics at mid-frequencies.
}
\label{fig: appendix-ExtremeLimits}
\end{figure*}

The model Hamiltonian Eq.~\eqref{eq: hamiltonian} becomes a free model when, e.g.~the anisotropy $\Delta$ is set to zero or, in the presence of $\delta \hat H_1$, for large $J_2 \rightarrow \infty$. In Sec.~\ref{sec: NearlyFree} we showed how approaching these free limits predicts the behaviour $\sigma'(\omega) \approx \omega^{-1}$ at the Ansatz level.
Here we corroborate this predictions.

First consider the small $\Delta$ limit in the XXZ chain, wherein the Ansatz simplifies to Eq.~\eqref{eq:smallDapprox_II}. 
We contrast in the left panel of Fig.~\ref{fig: appendix-ExtremeLimits} the Kubo conductivity between the isotropic point and the small $\Delta$ limit. 
Two points are noteworthy: (i) there are clear differences between the power laws about $\omega \approx 0.1$ for the two cases, with $\sigma(\omega) \sim \omega^{-1}$ being approximately satisfied for the small-$\Delta$ chain; 
(ii) also for $\Delta = 0.05$ there is a $\Delta^2$ drop between the end of the $\omega^{-1}$ power-law and the onset of the $\omega^2$ finite-size power law; this is 
in contrast to the isotropic point where the drop to the $\omega^2$ power law is immediate. Such a $\Delta^2$ drop is in line with the Ansatz prediction,  Eq.~\eqref{eq:smallDapprox_II}.

Now consider the large $J_2$ limit for which the Ansatz yields Eq.~\eqref{eq: largeJ2}. In this system the spin chain effectively decouples into two independent chains with weak inter-chain hopping.
We expect, from the results shown in the left panel of Fig.~\ref{fig: DiffusionExponent}, the $\pm J_2$ dynamical responses to be identical as $J_2 \rightarrow \infty$, with $\alpha \rightarrow -1$ (as in the small $\Delta$ limit treated above). 
This restoration of the symmetry and the actual power-law at mid-frequencies being close to -1 is supported by the numerics, as displayed in the right panel of Fig.~\ref{fig: appendix-ExtremeLimits}.

Let us nevertheless remark once again that for $1 \ll J_2 \neq \infty$ the system is still nonintegrable and therefore normal diffusion is expected at long times. 
%In this limit, for $\omega \ll \omega_0(J_2 \rightarrow \infty, 1) = \frac{\sqrt{5}}{2}$, our Ansatz predicts

%\begin{equation}
%\label{eq: largeJ2}
%\Sigma(\omega) \approx \frac{49 \pi}{10} \frac{1}{\omega^{1-\frac{49}{20J_2^2}}} + \mathcal{O}\left(\frac{1}{J_2}\right), \quad \mbox{for} \quad |J_2| \gg 1;
%\end{equation}
%
\section{Kubo conductivity: ensemble, boundary, and nnn-strength dependencies}
\label{app: conductivity}
 \label{app: FSEfluxing}
 
\subsection{Role of symmetries}
The infinite temperature conductivity at frequency $\omega$ is computed from the Kubo formula Eq.~\eqref{eq:Kubo}, which includes both the regular and Drude contributions. 

\begin{figure}[ttp!]
\centering 
\includegraphics[width=8.5cm]{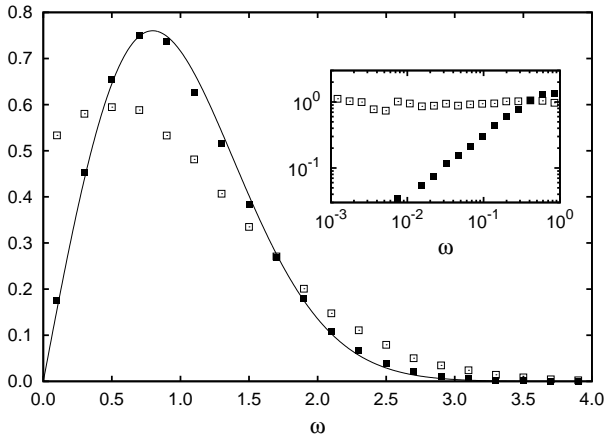}
\caption{Filled squares: level-spacing distribution for a 20-site spin ring with $\Delta = 1$ and $J_2=0.2$ averaged over the different symmetry subsectors with zero total magnetization ($m_z=0$). The inset shows the usual low-frequency linear decay of the level spacing distribution due to level repulsion. Empty squares: Level-spacing distribution for the states mixed by the spin current operator, Eq.~\eqref{eq:current}. Note the absence of level repulsion in this case which will be reflected in the conductivity. The ideal GOE distribution is shown as a full line in both plots.}
\label{fig: PDF}
\end{figure}

\vspace{0.2cm}
{\it a. Regular component}. First, a few comments on practical aspects vis-\'{a}-vis implementation are in order. Given that $\hat m_z$ and $\hat Z$ commute with both $\hat T$ and $\hat P$, and that $[\hat T, \hat P] = 0$ in the $k=0$ sector 
and $[\hat m_z, \hat Z] = 0$ in the $m_z=0$ sector, one can use this set of symmetries to block-diagonalize the Hamiltonian\citep{Sandvik}. 
For finite $J_2$ and large $L$ the energy-level statistics of each of these symmetry sectors are those of the Gaussian orthogonal ensemble (GOE)\citep{PhysRevLett.70.497}. 
However, as previously mentioned, the total spin current operator $\hat j^z$ is odd under both parity and spin-inversion and, consequently, 
its only nonvanishing matrix elements are those connecting symmetry subsectors with opposite $(p, z)$. 
Thus, in the presence of spin-inversion symmetry and parity, the energy difference in the argument of the Dirac delta functions in the Kubo formula mixes different GOE spectra. 
Fig.~\ref{fig: PDF} illustrates the consequences of these observations: the filled squares show the level statistics taken within each symmetry sector and then averaged over; 
the correspondence with random matrix theory is evident. 
Empty squares in Fig. \ref{fig: PDF}, in contrast, correspond to the level spacing distribution for the set of states mixed by the current operator in the $m_z = 0$ sector. 
The main aspect to be noted here is the absence of level repulsion, which entails that for frequencies much smaller than the average level spacing, the conductivity, when computed in the canonical ensemble within the zero-magnetization sector, will not display the usual reduction due to the decrease of level pairs with such energy differences.
This means that there is no discernible low-frequency $|\omega|$ drop in the conductivity, as can be clearly seen in the center panel of Fig.~\ref{fig: Conductivity_3} (wherein the average is done in a spin-inversion symmetric sector, i.e. with $m_z=0$). 

Now in the absence of parity and spin-inversion, all many-body states are mixed by the current operator and random matrix theory effectively describes the spacing distribution of energy levels.
This is the case upon the application of a flux: the flux breaks spin-inversion, parity and time-reversal symmetry (here labeled as $\hat K$) but preserves the symmetry under the combined operation $(\hat K \times \hat P)$. Level spacing statistic thus belongs to the GOE\citep{PhysRevLett.70.497}. 
Indeed, one can clearly identify the onset of level repulsion in the center and right panels of Fig.~\ref{fig:anomalyJ2} of the main text (which correspond to fluxed chains), as the linear drop of the conductivity at the lowest frequencies (marked with dashed lines). 

\vspace{0.2cm}
{\it b. Drude component}. As regards the ballistic contribution, the flux plays an important role in accelerating convergence specially in the integrable case\cite{SanchezVarma} --- this is due to the presence of additional symmetries, partially identified as the $\textrm{sl}_2$-loop symmetry. In the absence of the flux the overwhelming contributors to the Drude weight are states from these $\textrm{sl}_2$-loop degenerate subspace of off-diagonal current matrix elements $j_{m,n}^z$, with $m \neq n$. How fast these degeneracies appear in our finite chains, however, depends on the anisotropy value.
Upon introduction of the flux all the $\textrm{sl}_2$-loop-related degeneracies disappear and the relevant current carrying states are the diagonal ones i.e.~$j_{m,n}^z \delta_{m,n}$. Convergence to the thermodynamic limit is faster in this case.
Nevertheless, these two types of contributions in the two situations (i.e.~fluxed vs. unfluxed) approach the same value in the thermodynamic limit, irrespective of the anisotropy $\Delta$. 
For $\Delta \geq 1$ or $J_2 \neq 0$ this thermodynamic limit is zero.

Now, all our {\it finite-size} spin chains have a finite Drude weight (even for finite $J_2$) in the symmetry sectors of {\it finite} $m_z$ --- if $m_z=0$ there are no current-carrying states due to spin-inversion!. 
Breaking spin-inversion with a flux yields a finite Drude weight even within the zero total magnetization sector, thereby enhancing the finite-size grand-canonical ensemble value of the Drude weight\citep{SanchezVarma}. 
It follows that in fluxed integrable chains the regular conductivity should drop at higher frequencies as compared to unfluxxed chains, for the optical sum rule to hold. 
Comparing the left panels of Fig.~\ref{fig: Conductivity_3} and Fig.~\ref{fig:anomalyJ2} shows this is indeed the case. 
%
%Most importantly, note from Fig.~\ref{fig:anomalyJ2} the $\omega^2$ behavior shows up for $\omega \lesssim 0.1$ and moves slowly to lower frequencies when increasing system size, which identifies it as a finite-size effect.

\subsection{Role of ensembles}
\begin{figure*}[ttp!]
\centering 
\includegraphics[width=18.5cm]{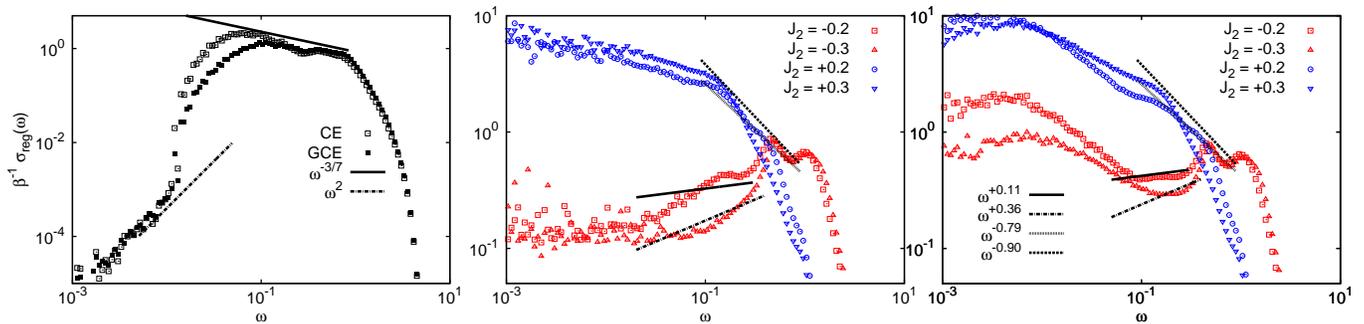}
\caption{Mid-frequency power laws and ensemble dependence of the (normalized) high temperature conductivity.
{\it Left}: conductivity in canonical (CE) and grand canonical ensemble (GCE) of the $L=18$ integrable chain.
A superdiffusive power-law as obtained from the ansatz gives a good-fit before the low-frequency $\omega^2$ behavior sets in; the latter is consistent with the argument in Ref. \cite{PhysRevB.86.115106} at 
low-frequencies in the integrable chains.
{\it Center}: conductivity in nonintegrable $L=18$ chains in the $m_z = 0$ sectors; the black lines correspond to power-law fits from the Ansatz, whose exponents are defined in the right panel of the figure. 
Note the behaviour predicted by the Ansatz is qualitatively and quantitatively consistent with the drift, and often values, of the power laws as $\pm J_2$ is changed.
{\it Right}: Same as middle panel but in the grand canonical ensemble. An extra peaking is observed at much lower-frequencies in all $J_2 \neq 0$ cases, 
which is a remnant of the finite-$L$ Drude peak from the integrable limit (which is absent in the middle panel due to spin-inversion symmetry).
}
\label{fig: Conductivity_3}
\end{figure*}

In Figs.~\ref{fig: DiffusionExponent} and~\ref{fig:anomalyJ2} we demonstrated the qualitative and quantitative dependence of low-frequency power-laws as a function of $J_2$, 
and their agreement with the numerical results.
To further study the long-time transport properties of our model, as well as the validity of the predictions let us consider Fig.~\ref{fig: Conductivity_3}: 
here we display the ensemble dependence of the conductivity, computed using logarithmically spaced frequency bins {\it in the absence of a flux}, together with the mid-frequency power-law predictions from the memory function Ansatz.
We distinguish between canonical (CE) and grand canonical (GCE) ensemble calculations: in the former case we average over eigenstates with $m_z=0$, whereas in the latter the average includes eigenstates 
from all the different magnetization sectors. Let us consider the integrable and nonintegrable case separately. 

\vspace{0.2cm}
{\it a. Integrable case}. The left panel of Fig.~\ref{fig: Conductivity_3} depicts the finite-size low-frequency conductivity for an 18-site isotropic Heisenberg chain. The first aspect ones notes is that at frequencies lower than $\omega\sim0.1$ the conductivity abruptly drops, a feature which has been identify as a pseudogap in previous studies\citep{PhysRevB.70.205129}. The difference between the two ensemble calculations at these frequencies has a simple explanation: at the isotropic point ($\Delta=1$) the Drude weight {\it of small chains} is finite and vanishes polynomially when increasing system size in the GCE, whereas it vanishes identically for any $L$ due to spin-inversion in the CE. 
Hence we expect the regular part of the GCE conductivity to drop at higher frequencies -- as compared with the CE calculation, in order to preserve the optical $f$ sum-rule fixed by $\mu_0$. 
Since in the thermodynamic limit the Drude weight vanishes, such a difference between ensembles is nothing but a finite size effect. At even lower frequencies we observe the onset of a clear $\omega^2$ power law; 
such a $\omega^2$ low-frequency behavior has been argued for in Ref.~\citen{PhysRevB.86.115106} for the integrable chain. 
However, as we have already seen in Fig.~\ref{fig:anomalyJ2} of the main text, the onset of the $\omega^2$ behavior slowly moves towards lower frequencies when increasing $L$, thus indicating it corresponds to another finite-size effect. 
Hence, for the system sizes considered here we expect strong finite-size effects for frequencies $\omega \lesssim 0.1$. 

Note that there is strong numerical evidence for superdiffusion at the isotropic point\cite{MarkoAnisotropic, PhysRevLett.117.040601, Ljubotina};
in the main text this was displayed in left panel of Fig.~\ref{fig:anomalyJ2}. 
The same superdiffusive power-law, as obtained from our Ansatz Eq.~\eqref{eq: ansatz}, nicely fits the CE data as well (as shown in left panel of Fig.~\ref{fig: Conductivity_3}) 
before the sudden drop of the conductivity. 

\vspace{0.2cm}
{\it b. Nonintegrable case}. 

Let us now consider the ensemble and $J_2$ dependence, focusing first on the center panel of Fig.~\ref{fig: Conductivity_3}, where we plotted CE data for four $J_2$ values. 
To begin with, note the strong suppression of spectral weight at low frequencies for negative $J_2$, and the absence of level repulsion at the lowest frequencies due to spin-inversion and parity symmetries 
(see Fig.~\ref{fig: PDF}). 
Most importantly, and focusing on frequencies $\omega \gtrsim 0.1$ (i.e.~before the finite-size effects observed in the integrable limit set in) we see there is a frequency range on 
which the power laws obtained from our Ansatz, represented as black lines whose legend is shown in the right panel of the figure, are qualitatively consistent with the drift and power laws of the conductivity 
data for both $\pm J_2$ cases. 
Pivotally one can clearly identify from the exact-diagonalization data alone power-law behavior which signals anomalous diffusion processes occurring on a limited time domain. 

%These observations are interesting because even if anomalous transport shows up on a limited frequency range only, it might be enough to generate anomalous behavior in NMR experiments, as was argued by 
%Sirker et al.~for the $\Delta<1$ regime of the integrable model\citep{PhysRevB.83.035115}: although normal diffusion was found to take place on a limited space-time domain, it was enough to give 
%diffusive behavior to spin-lattice relaxation rates, in agreement with experimental findings in the 1D chain $\textrm{Sr}_2\textrm{CuO}_3$. 
%%
The right panel of Fig.~\ref{fig: Conductivity_3} shows similar results for the conductivity computed in the GCE (system size dependence was already displayed in the main text for a given $J_2$ in the GCE). 
For $\omega \gtrsim 0.1$ the qualitative agreement between the data and the Ansatz predictions persists. 
At frequencies $\omega \lesssim 0.1$ however, the conductivity displays an additional peak which could be interpreted as follows: 
for finite $L$ the Drude weight in the integrable limit is finite; breaking integrability transfers the Drude peak's spectral weight to finite frequencies so that the optical sum-rule is 
satisfied\citep{PhysRevB.77.161101}. One might argue that such a redistribution results in the low-frequency enhancement at $\omega \lesssim 0.1$, which then corresponds to a finite-size effect. 

A bit more tricky is the increase with system size of the low-frequency enhancement before the onset of level repulsion --- as shown in Fig.~\ref{fig:anomalyJ2} of the main text. 
One can identify two competing effects: first, and as pointed out above, breaking integrability transfers the Drude peak's spectral weight {\it from the integrable limit} to finite frequencies; 
the strength of this effect should decrease when increasing system size. 
Second, the finite Drude weight of the nonintegrable chain itself rapidly decreases with system size, the corresponding spectral weight being shifted to finite frequencies. 
The combination of these two effects yields the low-frequency enhancement of the conductivity for $\omega \lesssim 0.1$, which then corresponds to a finite-size effect.

In summary, we have seen that the power laws obtained from the Ansatz, Eq.~\eqref{eq: ansatz}, qualitatively describe the finite-frequency conductivity power laws observed in both canonical and 
grand canonical ensemble computation. These results entail the possibility of mid-frequency anomalous diffusion in nonintegrable systems.

\bibliographystyle{unsrt}
\bibliography{Ref6}

\begin{thebibliography}{10}

\bibitem{ChaikinLubensky}
P.~M. Chaikin and T.~C. Lubensky.
\newblock {\em Principles of condensed matter physics}.
\newblock Cambridge University Press, 1995.

\bibitem{Mahan}
G.~D. Mahan.
\newblock {\em Many-particle Physics}.
\newblock Kluwer Academic / Plenum Publishers, New York, 2000.

\bibitem{PhysRevLett.74.972}
H.~Castella, X.~Zotos, and P.~Prelov\ifmmode~\check{s}\else \v{s}\fi{}ek.
\newblock {\em Phys. Rev. Lett.}, 74:972--975, 1995.

\bibitem{PhysRevB.53.983}
X.~Zotos and P.~Prelov\ifmmode~\check{s}\else \v{s}\fi{}ek.
\newblock {\em Phys. Rev. B}, 53:983--986, 1996.

\bibitem{PhysRevB.55.11029}
X.~Zotos, F.~Naef, and P.~Prelovsek.
\newblock {\em Phys. Rev. B}, 55:11029--11032, 1997.

\bibitem{continuum}
Models defined in the continuum generally have finite Drude weight if momentum
  conservation implied current conservation.

\bibitem{SanchezVarma}
R.~J. S\'anchez and V.~K. Varma.
\newblock {\em arXiv: 1704.04273}, 2017.

\bibitem{PhysRevB.77.161101}
M.~Rigol and B.~S. Shastry.
\newblock {\em Phys. Rev. B}, 77:161101, 2008.

\bibitem{PhysRevB.73.035113}
S.~Mukerjee, V.~Oganesyan, and D.~A. Huse.
\newblock {\em Phys. Rev. B}, 73:035113, 2006.

\bibitem{PhysRevA.92.013618}
V.~K. Varma and R.~J. S\'anchez.
\newblock {\em Phys. Rev. A}, 92:013618, 2015.

\bibitem{Mazur_ineq}
P~Mazur.
\newblock {\em Physica}, 43:533, 1969.

\bibitem{PhysRevLett.106.217206}
T.~Prosen.
\newblock {\em Phys. Rev. Lett.}, 106:217206, 2011.

\bibitem{PhysRevLett.111.057203}
T.~Prosen and E.~Ilievski.
\newblock {\em Phys. Rev. Lett.}, 111:057203, 2013.

\bibitem{DeguchiFabriciusMcCoy}
T.~Deguchi, K.~Fabricius, and McCoy~B. M.
\newblock {\em J. Stat. Phys.}, 102:701, 2001.

\bibitem{PhysRevB.92.165133}
J.~M.~P. Carmelo, T.~Prosen, and D.~K. Campbell.
\newblock {\em Phys. Rev. B}, 92:165133, 2015.

\bibitem{CarmeloProsen_UpperBoundXXXmodelGranCan}
J.~M.~P. Carmelo, T.~Prosen, and D.~K. Campbell.
\newblock {\em Nucl. Phys. B}, 914:62, 2017.

\bibitem{PhysRevLett.119.020602}
E.~Ilievski and J~De~Nardis.
\newblock {\em Phys. Rev. Lett.}, 119:020602, 2017.

\bibitem{PhysRevB.68.134436}
F.~Heidrich-Meisner, A.~Honecker, D.~C. Cabra, and W.~Brenig.
\newblock {\em Phys. Rev. B}, 68:134436, 2003.

\bibitem{ansatz1f}
We want to stress that, as it stands, Eq.~\eqref{eq: ansatz} assumes the spin
  current does not overlap with any of the local or quasilocal conserved
  quantities of the integrable model, i.e. no Drude weight, and hence it needs
  to be modified when applied to the $\Delta < 1$ regime of the XXZ model, as
  we do subsequently in the main text.

\bibitem{ansatz2f}
We choose a Gaussian Ansatz because it leads to non-infinite moments of any
  order, and allows to analytically solve the sum-rule equations for the
  parameters determining the Ansatz. Similar ideas have been used before in the
  context of high-temperature spin transport to obtain the diffusion constant
  of the integrable model\citep{deGennes, Mori, PhysRev.138.A608}.
  Eq.~\eqref{eq: ansatz} simply extends these ideas to allow for anomalous
  diffusion. Indeed for $\alpha = 0$ the Ansatz gives pure Gaussian decay,
  corresponding to diffusive modes i.e. $C(q,t) = \sum_j e^{i q \cdot (r_j -
  r_0)} ( S^{z}_j | S^{z}_0(t)) \sim e^{-\mathcal{D} \, q^2t}$, for the
  wavevector $q$ and spin diffusion constant $\mathcal{D}$. When $\alpha > 0$
  this decay is further suppressed with a higher power of $q$ and subdiffusion
  ensues; for $\alpha < 0$ transport becomes superdiffusive.

\bibitem{gapped}
We will avoid often used terminology of "gapped" and "gapless" phases which is
  meaningless near $T=\infty$.

\bibitem{ProsenGapless}
T.~Prosen.
\newblock {\em Phys. Rev. Lett.}, 106:217206, 2011.

\bibitem{Zotos}
X.~Zotos.
\newblock {\em Phys. Rev. Lett.}, 82:1764, 1999.

\bibitem{Benz}
J.~Benz, T.~Fukui, A.~Kluemper, and C.~Scheeren.
\newblock {\em J. Phys. Soc. Jpn. Supp.}, 74:181, 2005.

\bibitem{artefact}
We point that this smoothly varying exponent might be an artefact of our
  3-parameter ansatz. An obvious way to improve upon this is to introduce more
  sophisticated Ans\"{a}tze that require higher moments; so far our extended
  choices failed to provide a consistent hydrodynamic picture for the different
  systems we studied, i.e. systems with integrable and nonintegrable limits. We
  believe finding a judicious choice that captures the
  ballistic-superdiffusive-diffusive sharp jumps across the isotropic point, as
  numerics seems to suggest (although still only arguably) will not be easy,
  which we leave for future work.

\bibitem{PhysRevLett.119.080602}
Marko Medenjak, Christoph Karrasch, and Toma\ifmmode \check{z}\else~\v{z}\fi{}
  Prosen.
\newblock {\em Phys. Rev. Lett.}, 119:080602, 2017.

\bibitem{PhysRevB.49.3592}
M.~Chertkov and I.~Kolokolov.
\newblock {\em Phys. Rev. B}, 49:3592--3595, 1994.

\bibitem{PhysRevB.49.15669}
M.~B\"ohm, V.~S. Viswanath, J.~Stolze, and G.~M\"uller.
\newblock {\em Phys. Rev. B}, 49:15669--15681, 1994.

\bibitem{PhysRevB.57.8340}
K.~Fabricius and B.~M. McCoy.
\newblock {\em Phys. Rev. B}, 57:8340--8347, 1998.

\bibitem{MarkoAnisotropic}
M.~Znidaric.
\newblock {\em Phys. Rev. Lett.}, 106:220601, 2011.

\bibitem{PhysRevLett.117.040601}
M.~\ifmmode \check{Z}\else \v{Z}\fi{}nidari\ifmmode~\check{c}\else \v{c}\fi{},
  A.~Scardicchio, and V.~K. Varma.
\newblock {\em Phys. Rev. Lett.}, 117:040601, 2016.

\bibitem{Ljubotina}
M.~Ljubotina, M.~Znidaric, and T.~Prosen.
\newblock {\em arXiv: 1702.04210}, 2017.

\bibitem{deGennes}
P.~G. De~Gennes.
\newblock {\em J. Phys. Chem. Solids}, 4:223, 1958.

\bibitem{Mori}
H.~Mori and K.~Kawasaki.
\newblock {\em Prog. Theo. Phys.}, 27:529, 1962.

\bibitem{PhysRev.138.A608}
H.~S. Bennett and P.~C. Martin.
\newblock {\em Phys. Rev.}, 138:A608, 1965.

\bibitem{Auerbach}
I.~Khait, S.~Gazit, N.~Y. Yao, and A.~Auerbach.
\newblock {\em Phys. Rev. B}, 93:224205, 2016.

\bibitem{PhysRevLett.96.067202}
P.~Jung, R.~W. Helmes, and A.~Rosch.
\newblock {\em Phys. Rev. Lett.}, 96:067202, 2006.

\bibitem{PhysRevB.76.245108}
Peter Jung and Achim Rosch.
\newblock {\em Phys. Rev. B}, 76:245108, 2007.

\bibitem{Roldan}
J.~M.~R. Roldan, B.~M. McCoy, and J.~H.H. Perk.
\newblock {\em Physica A: Statistical Mechanics and its Applications}, 136:255,
  1986.

\bibitem{Steinigeweg}
R.~Steinigeweg and R.~Schnalle.
\newblock {\em Phys. Rev. E}, 82:040103, 2010.

\bibitem{Karrasch}
C.~Karrasch, J.~E. Moore, and F.~Heidrich-Meisner.
\newblock {\em Phys. Rev. B}, 89:075139, 2014.

\bibitem{Dorfman1970}
J.~R. Dorfman and E.~G.~D. Cohen.
\newblock {\em Phys. Rev. Lett.}, 25:1257, 1970.

\bibitem{PhysRevB.77.245131}
S~Mukerjee and B.~S. Shastry.
\newblock {\em Phys. Rev. B}, 77:245131, 2008.

\bibitem{Katsura}
S.~Katsura, T.~Horiguchi, and M.~Suzuki.
\newblock {\em Physica}, 46:67, 1970.

\bibitem{McCoySpin}
B.~M. McCoy.
\newblock {\em Phys. Rev.}, 173:531, 1968.

\bibitem{dispersion}
There are two set of results that establish the structure of phonon-like
  features in the dynamic structure factor. At $T=0$ recent works (e.g. by
  Imambekov et al \cite{Glazman}) established that Luttinger bosons fragment
  into a narrow band centered about $v_F q$ of width $\sim q^3$. There is no
  central peak at T=0. Near infinite temperature Ref.
  \citen{PhysRevB.77.245131} presented clear numerical evidence for the
  existence of a dispersive feature in the momentum dependent conductivity
  which is related to dynamical structure factor via $(\omega/q)^2$. For XY
  chains closed form analytic results are available \cite{Katsura}, with
  dynamical structure factor displaying threshold (square root) singularities
  at $\pm v_F q$.

\bibitem{Doyon}
O.~A. Castro-Alvaredo, B.~Doyon, and T.~Yoshimura.
\newblock {\em Phys. Rev. X}, 6:041065, 2016.

\bibitem{PhysRevLett.117.207201}
Bruno Bertini, Mario Collura, Jacopo De~Nardis, and Maurizio Fagotti.
\newblock {\em Phys. Rev. Lett.}, 117:207201, 2016.

\bibitem{PhysRevB.96.115124}
Lorenzo Piroli, Jacopo De~Nardis, Mario Collura, Bruno Bertini, and Maurizio
  Fagotti.
\newblock {\em Phys. Rev. B}, 96:115124, 2017.

\bibitem{PhysRevB.97.045407}
Vir~B. Bulchandani, Romain Vasseur, Christoph Karrasch, and Joel~E. Moore.
\newblock {\em Phys. Rev. B}, 97:045407, 2018.

\bibitem{DiracQ}
J.G. Wright and B.~S. Shastry.
\newblock {\em DiracQ: A Quantum Many-Body Physics Package}.
\newblock arXiv:1301.4494 [cond-mat,str-el], 2013.

\bibitem{Berne}
B.~J. Berne and G.~D. Harp.
\newblock {\em Adv. Chem. Phys.}, XVII:63, 1970.

\bibitem{Forster}
D.~Forster.
\newblock {\em Hydrodynamic fluctuations, broken symmetry and correlation
  functions}.
\newblock W. A. Benjamin, Inc., 1975.

\bibitem{Sandvik}
A.~W. Sandvik.
\newblock {\em AIP Conf. Proc}, 1297:135, 2010.

\bibitem{PhysRevLett.70.497}
G.~Montambaux, D.~Poilblanc, J.~Bellissard, and C.~Sire.
\newblock {\em Phys. Rev. Lett.}, 70:497, 1993.

\bibitem{PhysRevB.86.115106}
J.~Herbrych, R.~Steinigeweg, and P.~Prelov\ifmmode~\check{s}\else \v{s}\fi{}ek.
\newblock {\em Phys. Rev. B}, 86:115106, 2012.

\bibitem{PhysRevB.70.205129}
P.~Prelov\ifmmode~\check{s}\else \v{s}\fi{}ek, S.~El Shawish, X.~Zotos, and
  M.~Long.
\newblock {\em Phys. Rev. B}, 70:205129, 2004.

\bibitem{Glazman}
A.~Imambekov, T.~L. Schmidt, and L.~I. Glazman.
\newblock {\em Rev. Mod. Phys.}, 84:1253, 2012.

\end{thebibliography}

\end{document}